\documentstyle[12pt]{article}

\let\ni=\noindent

\begin{document}

\baselineskip 0.75cm
 
\pagestyle {empty}

\renewcommand{\thefootnote}{\fnsymbol{footnote}}

\newcommand{\CKM}{Cabibbo---Kobayashi---Maskawa }

~~~
\begin{flushright}
IFT-97/21
\end{flushright}

\vspace{1.0cm}

{\large \centerline {\bf Proposal of unified fermion texture\footnote{Work 
supported in part by the Polish KBN--Grant 2--B302--143--06.}}}

\vspace{1.0cm}

{\centerline {\sc Wojciech Kr\'{o}likowski}}

\vspace{0.7cm}

{\centerline {\it Institute of Theoretical Physics, Warsaw University}}

{\centerline {\it Ho\.{z}a 69,~~PL--00--681 Warszawa, ~Poland}}

\vspace{1.0cm}

{\centerline {\bf Abstract}}

\vspace{0.3cm}

 A unified form of mass matrix is proposed for neutrinos, charged leptons, up 
quarks and down quarks. Some constraints for the parameters involved are tent%
atively postulated. Then, the predictions are neatly consistent with available 
experimental data. Among the predictions are: {\it (i)} $ m_\tau \simeq 1776.80
$~MeV (with the inputs of $ m_e $ and $ m_\mu $), {\it (ii)} $ m_{\nu_0} \ll 
m_{\nu_1} \sim (0.6\;\,{\rm to}\;\,4)\times 10^{-2}\,$eV and $ m_{\nu_2} \sim 
(0.2\;\,{\rm to}\;\,1)\times 10^{-1}\,$eV (with the atmospheric--neutrino inpu%
ts of $ |m_{\nu_2}^2 - m_{\nu_1}^2| \sim (0.0003\;\,{\rm to}\;\,0.01)\,{\rm eV}
^2 $ and {\it the $\nu_\mu \rightarrow \nu_\tau $ oscillation amplitude $\sim 
0.8 $}), and also ({\it iii}) $ m_s \simeq 270\;{\rm MeV}$, $|V_{ub}/V_{cb}| 
\simeq 0.082 $ and arg$V_{ub} \simeq -64^\circ $ (with the inputs of $ m_c = 
1.3 $ GeV, $ m_b = 4.5 $ GeV, $|V_{us}| = 0.221 $ and $|V_{cb}| = 0.041 $, 
where $ m_u \ll m_c \ll m_t $ and $ m_d \ll m_s \ll m_b $). All elements of 
the \CKM matrix are evaluated. All elements of its lepton counterpart are 
calculated up to an unknown phase (Appendix B). Some items related to dynamical
aspects of the proposed fermion "texture" are briefly commented on (Appendix 
A). In particular, the notion of a novel dark matter, free of any Standard--%
Model interactions (and their supersymmetric variants), appears in the case of 
preon option.

\vspace{0.3cm} 

\ni PACS numbers: 12.15.Ff , 12.90.+b 
 
\vspace{1.5cm} 

\ni December 1997

\vfill\eject

\pagestyle {plain}

\setcounter{page}{1}

~~~

\vspace{0.8cm}

\ni {\bf 1. Introduction}

\vspace{0.3cm}

 For the last few years we studied the "texture" of fermion mass matrices, sta%
rting from charged leptons $ e^-\,,\,\mu^-\,,\,\tau^- $ (the first and second 
Ref. [1]), then considering up and down quarks $ u\,,\,c\,,\,t $ and $ d\,,\,s
\,,\,b $ (the third Ref. [1]), and finally extending the argument to neutrinos 
(the fourth Ref. [1]). In consequence, we came to a proposal of common
structure of four mass matrices $\widehat{M}^{(\nu)}$ , $\widehat{M}^{(e)}$ , 
$\widehat{M}^{(u)}$ and $\widehat{M}^{(d)}$ in the three--dimensional family 
space of neutrinos $(\nu)$, charged leptons $(e)$, up quarks $(u)$ and down 
quarks $(d)$, respectively.

 Explicitly, we proposed that

\begin{equation}
\widehat{M}^{(f)} = \frac{1}{29} \left(\begin{array}{ccc} 
\mu^{(f)}\varepsilon^{(f)\,2} & 2\alpha^{(f)} e^{i\varphi^{(f)}} & 0 \\ & & 
\\ 2\alpha^{(f)} e^{-i\varphi^{(f)}} & 4\mu^{(f)}(80 + \varepsilon^{(f)\,2})/9 
& 8\sqrt{3}(\alpha^{(f)} - \beta^{(f)}) e^{i\varphi^{(f)}} \\ & & 
\\ 0 & 8\sqrt{3}(\alpha^{(f)} - \beta^{(f)}) e^{-i\varphi^{(f)}} & 24\mu^{(f)}
(624 + 25C^{(f)}+ \varepsilon^{(f)\,2})/25 \end{array}\right)\; ,
\end{equation}

\ni where $ f = \nu\,,\,e\,,\,u\,,\,d $, while $\mu^{(f)}$, $\varepsilon^{(f)
\,2}$, $\alpha^{(f)}$, $\beta^{(f)}$, $ C^{(f)}$ and $\varphi^{(f)}$ denoted
real constants to be determined from the experimental data for fermion masses 
and mixing parameters. The proposed form (1) followed from: {\it (i)} an
idea about the origin of three fermion families, and {\it (ii)} an ansatz for 
the fermion mass matrix expressed in terms of the suggested family character%
istics. Note that the mass matrices $\widehat{M}^{(e)} $, $\widehat{M}^{(u)}$,
$\widehat{M}^{(d)} $ as given in Eq. (1) do not take the popular Georgi--%
Jarlskog form [2] for any choice of their parameters, if $\mu^{(f)} > 0 $.

 In the present paper, we do not go systematically into any motivation for the 
proposal (1), considering it simply as a detailed conjecture. The interested 
Reader may look for roots of the formula (1) in Refs. [1] (note that in Refs. 
[3] there was discussed a mass formula a bit different, especially in the 
quark case).

 Instead, we proceed in the present paper a step further with our conjecture 
(1), postulating tentatively the following constraints for the parameters 
$\alpha^{(f)}\;,\;\beta^{(f)}$ and $ C^{(f)}$ appearing there:

\vfill\eject

\vspace{-0.4cm}

\begin{eqnarray}
\alpha^{(\nu)}:\alpha^{(e)} & = & |Q^{(\nu)}|:|Q^{(e)}|\;, \nonumber \\
\alpha^{(u)}:\alpha^{(d)} & = & |Q^{(u)}|:|Q^{(d)}|\;, \nonumber \\
\beta^{(\nu)} & = & \beta^{(e)} \;, \nonumber \\
\beta^{(u)} & = & \beta^{(d)} \;, \nonumber \\
\left(\beta^{(\nu)} + \beta^{(e)}\right) : \left(\alpha^{(\nu)} + \alpha^{(
e)} \right) & = & \delta^{(l)}:\left(|Q^{(\nu)}| + |Q^{(e)}|\right)  \;, 
\nonumber \\
\left(\beta^{(u)} + \beta^{(d)}\right) : \left(\alpha^{(u)} + \alpha^{(d)} 
\right) & = & \left(B^{(q)} + \delta^{(q)} \frac{1}{N_C^{(q)} }\right) :
\left(|Q^{(u)}| + |Q^{(d)}|\right) \;, \nonumber \\
C^{(\nu)} = C^{(e)} = 0 & , & C^{(u)} > 0 \;,\; C^{(d)} = 0 \;.
\end{eqnarray}

\vspace{-0.1cm}

\ni Here, $ Q^{(\nu)} = 0 $, $ Q^{(e)} = -1 $, $ Q^{(u)} = 2/3 $, $ Q^{(d)}
= -1/3 $, $ B^{(q)} = 1/3 $ and $ N_C^{(q)} = 3 $, while $ 0 < \delta^{(l)} \ll
1 $ and  $ 0 < \delta^{(q)} \ll 1 $ (we may introduce into Eqs. (2) also $ B^{(
l)} = 0 $ and $ N_C^{(l)} = 1 $, making then these relations fully symmetric 
under the interchange of leptons and quarks, if $ C^{(f)}$ are treated there 
like some charges). Thus, from Eqs. (2)

\vspace{-0.3cm}

\begin{eqnarray*}
\alpha^{(\nu)} = 0\; & , &\;\alpha^{(u)} = 2\alpha^{(d)}> 0\;,\\
\beta^{(\nu)} = \beta^{(e)} = \delta^{(l)}\alpha^{(e)}/2 \stackrel{>}{\sim} 0 
& , & \beta^{(u)} = \beta^{(d)} = (1 + \delta^{(q)})\alpha^{(d)}/2 \simeq 
\alpha^{(d)}/2 > 0 
\end{eqnarray*}

\vspace{-0.1cm}

\ni (we choose $\alpha^{(e)} > 0 $ and $\alpha^{(d)} > 0 $). We also assume 
that $\mu^{(\nu)} \simeq 0 $ and $\varepsilon^{(\nu)\,2} \simeq 0 $.

 Then, four mass matrices (1) contain practically 14 independent parameters, 
say,

\vspace{-0.3cm}

$$
\mu^{(\nu)} \simeq 0\;\,,\,\;\mu^{(e)}\;\,,\,\;\varepsilon^{(\nu)\,2} \simeq 
0\,\;,\;\,\varepsilon^{(e)\,2} \;\,,\,\; \beta^{(\nu)} \simeq 0\;\,,\,\;
\alpha^{(e)}\;\,,\,\;\varphi^{(\nu)} - \varphi^{(e)}
$$

\vspace{-0.1cm}

\ni and

\vspace{-0.3cm}

$$
\mu^{(u)}\;\,,\,\;\mu^{(d)}\;\,,\,\;\varepsilon^{(u)\,2}\,\;,\;\,
\varepsilon^{(d)\,2}\;\,,\,\;C^{(u)}\;\,,\,\;\alpha^{(d)}\;\,,\,\;\varphi^{(u)}
- \varphi^{(d)}\;,
$$

\vspace{-0.1cm}

\ni 7 for leptons and 7 for quarks (in addition, they contain $\varphi^{(\nu)} 
+ \varphi^{(e)}$ and $\varphi^{(u)} + \varphi^{(d)}$ that, however, will not 
appear in experimentally measured quantities). These 14 free parameters will 
describe 12 fermion masses and their 8 mixing parameters, together 20 quanti%
ties, of which 14 may be used as inputs determining consistently all parameters
(except for $\varphi^{(\nu)} + \varphi^{(e)}$ and $\varphi^{(u)} + \varphi^{(d)
}$ that will remain undetermined, but may be put zero as being physically 
irrelevant). So, we will be able to get $ 20 - 14 = 6 $ predictions, 3 for 
leptons and 3 for quarks, and also an overall consistent determination of all 
parameters (except for the unphysical two). The agreement with available 
experimental data will turn out to be satisfactory.

 In the framework of mass matrices (1), the observed differences between spect%
ral properties of four types of fermions $ f = \nu\;,\;e\;,\;u\;,\;d $ will 
follow (in a large extent) from the interplay of magnitudes of the parameters 
$\mu^{(f)}$ contained in the diagonal elements of $\widehat{M}^{(f)}$ and the 
parameters $\alpha^{(f)}$ and $\beta^{(f)}$ appearing in its off--diagonal 
elements. Their ratios, $\alpha^{(f)}/\mu^{(f)}$ and $\beta^{(f)}/\mu^{(f)}$, 
will play the role of coupling constants in our "texture dynamics" ({\it cf.} 
Appendix A).

\vspace{0.3cm}

\ni {\bf 2. Charged leptons}

\vspace{0.3cm}

 In the case of charged leptons, we will assume that the off--diagonal elements
of the mass matrix $\widehat{M}^{(e)} = \left(M_{ij}^{(e)}\right)\;\;(i,j = 0,
1,2)$ given in Eq. (1) can be treated as a small perturbation of the diagonal 
terms. Then, in the lowest (quadratic) perturbative order we obtain

\vspace{-0.3cm}

\begin{eqnarray}
m_e & = & \frac{\mu^{(e)}}{29} \left[\varepsilon^{(e)\,2} - \frac{36}{320 - 5
\varepsilon^{(e)\,2}}\left(\frac{\alpha^{(e)}}{\mu^{(e)}}\right)^2\right]\; , 
\nonumber \\ 
m_\mu & = & \frac{\mu^{(e)}}{29} \left[\frac{4}{9}\left(80 + \varepsilon^{(e)
\,2}\right) + \frac{36}{320 - 5\varepsilon^{(e)\,2}}\left(\frac{\alpha^{(e)}
}{\mu^{(e)}}\right)^2 - \frac{10800}{31696 + 29\varepsilon^{(e)\,2}}\left(
\frac{\alpha^{(e)} - \beta^{(e)}}{\mu^{(e)}}\right)^2\right] \; ,\nonumber \\ 
m_\tau & = & \frac{\mu^{(e)}}{29} \left[\frac{24}{25}\left(624 + \varepsilon^{(
e)\,2}\right) + \frac{10800}{31696 + 29\varepsilon^{(e)\,2}}\left(
\frac{\alpha^{(e)} - \beta^{(e)}}{\mu^{(e)}}
\right)^2\right] \;\;.
\end{eqnarray}

\vspace{-0.1cm}

\ni These mass formulae give

\vspace{-0.3cm}

\begin{eqnarray}
m_\tau & = & 1776.80\;\; {\rm MeV} \nonumber \\ 
& + & \frac{216 \mu^{(e)}}{3625} \left[ \frac{111550}{31696 + 29
\varepsilon^{(e)\,2}}\,\left(\frac{\alpha^{(e)} - \beta^{(e)}}{\mu^{(e)}}
\right)^2 - \frac{487}{320 - 5\varepsilon^{(e)\,2}}\left(\frac{\alpha^{(e
)}}{\mu^{(e)}}\right)^2\right] \;,\nonumber \\
\varepsilon^{(e)\,2} & = & 0.172329 + O\left[\left(\frac{\alpha^{(e)}}{\mu^{(e)
}}\right)^2 \right] \;,\nonumber \\
\mu^{(e)} & = & 85.9924\;{\rm MeV} + O\left[\left(\frac{\alpha^{(e)}}{\mu^{(e)}
}\right)^2\right]\mu^{(e)} + O\left[\left(\frac{\alpha^{(e)} - \beta^{(e)}}{
\mu^{(e)}}\right)^2\right]\mu^{(e)}\;,
\end{eqnarray}

\vspace{-0.1cm}

\ni when the experimental values of $ m_e $ and $ m_\mu $ [4] are used as 
inputs. Then, in the first Eq.~(4) $\;6(351 m_\mu - 136 m_e)/125 = 1776.80
$~MeV, in the second $\; 320 m_e/(9 m_\mu - 4 m_e) = $ $ 0.172329 $ and in the 
third $\; 29(9 m_\mu - 4 m_e)/320 = 85.9924$ MeV.

 With $\beta^{(e)}$ neglected {\it versus} $\alpha^{(e)}$ due to Eq. (2), the 
first Eq. (4) gives

\vspace{-0.2cm}

\begin{equation}
m_\tau = \left[1776.80 + 10.2112\,\left(\frac{\alpha^{(e)}}{\mu^{(e)}}\right)^2
\right]\;{\rm MeV} \;,
\end{equation}

\vspace{-0.1cm}

\ni what shows that

\vspace{-0.2cm}
							
\begin{equation}
\left(\frac{\alpha^{(e)}}{\mu^{(e)}}\right)^2 = 0.020^{+0.029}_{-0.020} \;,
\end{equation}

\vspace{-0.1cm}

\ni when the experimental value $m_\tau = 1777.00^{+0.30}_{-0.27} $ MeV [4] is 
used as another input. Thus, as yet, the values of $\alpha^{(f)}$ and $\beta^{
(f)}$ are consistent with zero. We can see that the mass--matrix formula (1) 
predicts excellently the mass $ m_\tau $, even in the zero--order perturbative
calculation [1].

 The unitary matrix $\widehat{U}^{(e)}$, diagonalizing the mass matrix $
\widehat{M}^{(e)}$ according to the equality $\widehat{U}^{(e)\,\dagger}
\widehat{M}^{(e)}\widehat{U}^{(e)} = $ diag($ m_e\,,\,m_\mu\,,\,m_\tau $),
gets in the lowest (linear) perturbative order the form

\vspace{-0.2cm}

\begin{equation}
\widehat{U}^{(e)} = \widehat{1} + \frac{1}{29}\left(\begin{array}{ccc} 
0 & 2\frac{\alpha^{(e)}}{m_\mu}e^{i\varphi^{(e)}} & 0 \\ - 2\frac{\alpha^{(e)}}
{m_\mu} e^{-i\varphi^{(e)}} & 0 & 8\sqrt{3}\frac{
\alpha^{(e)} - \beta^{(e)}}{m_\tau}e^{i\varphi^{(e)}} \\ 0 & -8\sqrt{3}\frac{
\alpha^{(e)} - \beta^{(e)}}{m_\tau}e^{-i\varphi^{(e)}} & 0 \end{array} 
\right) \;,
\end{equation}

\vspace{-0.1cm}

\ni where the small $\varepsilon^{(e)\,2}$ is neglected. Here, due to Eq. (2),
$\beta^{(e)}$ can be also neglected {\it versus} $\alpha^{(e)}$.

\vspace{0.3cm}

\ni {\bf 3. Neutrinos}

\vspace{0.3cm}

 In the case of neutrinos, the mass matrix $\widehat{M}^{(\nu)} = \left(M^{(\nu
)}_{ij} \right)\;\;(i,j = 0,1,2) $ as given in Eq. (1), with $\alpha^{(\nu)} 
= 0 $ due to Eq. (2), leads exactly to the following eigen\-values interpreted
as neutrino masses:

\vspace{-0.2cm}

\begin{eqnarray}
m_{\nu_0}\;\;\; & = & M^{(\nu)}_{00} = \frac{\mu^{(\nu)}}{29}\varepsilon^{(\nu
)\,2} \;, \nonumber \\ 
m_{\nu_1,\nu_2} & = & \frac{M_{11}^{(\nu)} + M_{22}^{(\nu)}}{2} \mp \left[
\left( \frac{M_{11}^{(\nu)} - M_{22}^{(\nu)}}{2}\right)^2 +|M_{12}^{(\nu)} |^2 
\right]^{1/2} \nonumber \\ 
& = & \left[ 10.9 \mp 0.478 \frac{\beta^{(\nu)}}{\mu^{(\nu)}} \sqrt{1 + \left(
20.3 \frac{\mu^{(\nu)}}{\beta^{(\nu)}} \right)^2 }\right] \mu^{(\nu)}\;,
\end{eqnarray}

\ni where in~~$m_{\nu_1}$~~and~~$m_{\nu_2}$~~the very small~~$\varepsilon^{(
\nu)\,2}$~~is neglected in the second step.~~Here, $\left(M^{(\nu)}_{11} - m_i
\right)\left(M^{(\nu)}_{22} - m_i\right) = |M^{(\nu)}_{12}|^2\;\;(i = 1,2) $.

 The corresponding unitary matrix $\widehat{U}^{(\nu)}$ diagonalizing the mass 
matrix $\widehat{M}^{(\nu)}$ according to the equality $\widehat{U}^{(\nu)\,
\dagger}\widehat{M}^{(\nu)}\widehat{U}^{(\nu)} = $ diag($m_{\nu_0}\,,\,m_{
\nu_1}\,,\,m_{\nu_2}$), takes exactly the form

\begin{equation}
\widehat{U}^{(\nu)} = \left(\begin{array}{ccc} 1 & 0 & 0 \\ & & \\ 0 & (1 + X^2)^{-
1/2} & X(1 + X^2)^{-1/2} e^{i\varphi^{(\nu)}} \\ & & \\ 0 & - X(1 + X^2)^{-1/2}
e^{-i\varphi^{(\nu)}} & (1 + X^2)^{-1/2} \end{array}\right)\;,
\end{equation}

\vspace{0.2cm}

\ni where 

\begin{eqnarray}
X & = & \frac{M_{11}^{(\nu)} - m_{\nu_1}}{|M_{12}^{(\nu)}|} = \frac{M_{22}^{(
\nu)} - m_{\nu_2}}{|M_{12}^{(\nu)}|} \nonumber \\
& = & \frac{M_{11}^{(\nu)} - M_{22}^{(\nu)}}{2 |M_{12}^{(\nu)}|} + \left[1 + 
\left( \frac{M_{11}^{(\nu)} - M_{22}^{(\nu)}}{2 |M_{12}^{(\nu)} |}\right)^2
\right]^{1/2} \nonumber \\
& = & - 20.3 \frac{\mu^{(\nu)}}{\beta^{(\nu)}} + \sqrt{1 + \left(20.3 
\frac{\mu^{(\nu)}}{\beta^{(\nu)}} \right)^2}\;.
\end{eqnarray}

\vspace{0.2cm}

 The experimentally observed neutrino weak--interaction states $\nu_e\,,\,
\nu_\mu\,,\,\nu_\tau $ are related to their mass states $\nu^{\rm (m)}_0\,,\,
\nu^{\rm (m)}_1\,,\,\nu^{\rm (m)}_2 $ (corresponding to the masses $m_{\nu_0}\,,\,
m_{\nu_1}\,,\,m_{\nu_2}$) through the unitary transformation

\begin{eqnarray}
\left(\begin{array}{c}\nu_e  \\ \nu_\mu \\ \nu_\tau \end{array} \right)
= \widehat{V}^\dagger \left(\begin{array}{c}\nu_0^{\rm (m)} \\ \nu_1^{\rm (m)} \\ 
\nu_2^{\rm (m)} \end{array} \right)\, ,
\end{eqnarray}

\ni where 

\begin{equation}
\widehat{V} = \widehat{U}^{(\nu)\,\dagger} \widehat{U}^{(e)}
\end{equation}

\vspace{0.2cm}

\ni is the lepton \CKM matrix. Making use of Eqs. (9) and (7), we can calculate
$\widehat{V} = \left( V_{ij}\right)\;\; (i,j = 0,1,2)$ from the formulae $
V_{ij} = \sum_k{U_{ki}^{(\nu)\,*}U_{kj}^{(e)}}$. The result, valid in the 
lowest (linear) perturbative order in $\alpha^{(e)}/\mu^{(e)}$ and 
$\beta^{(e)}/\mu^{(e)}$, reads

\vfill\eject

\begin{eqnarray}
V_{01} & = & \frac{2}{29}\frac{\alpha^{(e)}}{m_\mu}e^{i\varphi^{(e)}}\;,\; 
V_{10} = -\frac{2}{29\sqrt{1+X^2}}\frac{\alpha^{(e)}}{m_\mu}
e^{-i\varphi^{(e)}}\;, \nonumber \\  V_{12} & = & -\frac{X}{\sqrt{1+X^2}}
e^{i\varphi^{(\nu)}} + \frac{8\sqrt{3}}{29\sqrt{1+X^2}}\frac{\alpha^{(e)} - 
\beta^{(e)}}{m_\tau}e^{i\varphi^{(e)}} = - V^*_{21}\;, \nonumber \\  
V_{02} & = & 0\;,\; V_{20} = -\frac{2X}{29\sqrt{1+X^2}}\frac{\alpha^{(e)}}
{m_\mu}e^{-i(\varphi^{(\nu)}+\varphi^{(e)})}\;, \nonumber \\  V_{00} & = & 1
\;,\; V_{11} = \frac{1}{\sqrt{1+X^2}} + \frac{8\sqrt{3}X}{29\sqrt{1+X^2}}
\frac{\alpha^{(e)}-\beta^{(e)}}{m_\tau}e^{i(\varphi^{(\nu)}-\varphi^{(e)})} 
=  V^{*}_{22}\;,
\end{eqnarray}

\ni where $\beta^{(e)}$ can be neglected {\it versus} $\alpha^{(e)}$ (for the
numerical form of $ V_{ij}$ {\it cf.} Appendix B).

 We can see from Eqs (11) and (13) that

\begin{eqnarray}
\nu^{\rm (m)}_0 & = & \nu_e + O\left(\frac{\alpha^{(e)}}{m_\mu}\right) \nu_\mu
\;, \nonumber \\ 
\nu^{\rm (m)}_1 & = & \left[\frac{1}{\sqrt{1+X^2}} + O\left(
\frac{\alpha^{(e)} - \beta^{(e)}}{m_\tau}\right)\right]\nu_\mu - 
\left[\frac{X}{\sqrt{1+X^2}}e^{i\varphi^{(\nu)}} + O\left(\frac{\alpha^{(e)} -
\beta^{(e)}}{m_\tau}\right)\right]\nu_\tau \nonumber \\ & + & O\left(
\frac{\alpha^{(e)}}{m_\mu}\right) \nu_e \;, \nonumber \\ 
\nu^{\rm (m)}_2 & = & \left[\frac{X}{\sqrt{1+X^2}}e^{-i\varphi^{(\nu)}} + O\left(
\frac{\alpha^{(e)} - \beta^{(e)}}{m_\tau}\right)\right]\nu_\mu + \left[\frac{1}
{\sqrt{1+X^2}} + O\left(\frac{\alpha^{(e)} - \beta^{(e)}}{m_\tau}\right)\right]
\nu_\tau \nonumber \\ 
& + & O\left(\frac{\alpha^{(e)}}{m_\mu}\right) \nu_e \;.
\end{eqnarray}

\ni Thus, if $ X $ is of the order $ O(1) $, in Eqs. (14) there appears strong 
mixing between $\nu_\mu $ and $\nu_\tau $ beside weak mixing of $\nu_\mu $ 
and $\nu_\tau $ with $\nu_e $. Note from Eq. (10) that $ X \rightarrow 1 $ in 
the limit of $\mu^{(\nu)}/\beta^{(\nu)} \rightarrow 0 $ and, for example, $ X =
\sqrt{2} - 1 $ or $ (\sqrt{5} - 1)/2 $ for $\mu^{(\nu)}/\beta^{(\nu)} = 1/20.3$
or $ 1/40.6 $, respectively (in these cases $ 4X^2(1 + X^2)^{-2} \rightarrow 
1 $ and $ 4X^2(1 + X^2)^{-2} = 0.5 $ or 0.8).

 Once we know the elements (13) of the lepton \CKM matrix $\widehat{V}$, we can
calculate the probabilities of neutrino oscillations $\nu_i \rightarrow \nu_j $
(in the vacuum) from the familiar formulae

\vspace{-0.15cm}

\begin{equation}
P(\nu_i \rightarrow \nu_j,t) = |\langle\nu_j |\nu_i(t)\rangle|^2 = \sum_{k\,l}
V^*_{j\,l}V_{i\,l}V_{j\,k}V^*_{i\,k} \exp\left(i\frac{m^2_{\nu_l}-m^2_{\nu_k}}
{2|\vec{p}|}\,t\right)\;,
\end{equation}

\ni where usually $ t/|\vec{p}| = L/E $ (what is equal to $ 4\times 1.2663 L/E$
if $m^2_{\nu_l} - m^2_{\nu_k}$, $ L $ and $ E $ are measured in eV$^2$, km and
GeV, respectively). Here, $ L $ is the source--detector distance. In the notat%
ion of Eq. (15), $\nu_0 \equiv \nu_e\,,\,\nu_1 \equiv \nu_\mu\,,\,\nu_2 \equiv 
\nu_\tau $ are the neutrino weak--interaction states (to be distinguished from 
their mass states $\nu^{\rm (m)}_0\,,\,\nu^{\rm (m)}_1\,,\,\nu^{\rm (m)}_2 $ 
corresponding to $ m_{\nu_0}\,,\,m_{\nu_1}\,,\,m_{\nu_2} $).

 After some calculations, we obtain in the lowest (linear and quadratic) 
perturbative order in $\alpha^{(e)}/\mu^{(e)}$ and $\beta^{(e)}/\mu^{(e)}$ the 
following formulae:

\vspace{-0.15cm}

\begin{eqnarray}
P\left(\nu_e \rightarrow \nu_\mu,t \right) & = & \frac{16}{841(1+X^2)} 
\left(\frac{\alpha^{(e)}}{m_\mu}\right)^2 \sin^2 \left(\frac{m_{\nu_1}^2 - 
m_{\nu_0}^2}{4|\vec{p}|}\,t\right)\;, \nonumber \\
P\left(\nu_e \rightarrow \nu_\tau,t \right) & = & \frac{16 X^2}{841(1+X^2)} 
\left(\frac{\alpha^{(e)}}{m_\mu}\right)^2 \sin^2 \left(\frac{m_{\nu_1}^2 - 
m_{\nu_0}^2}{4|\vec{p}|}\,t\right)\;, \nonumber \\
P\left(\nu_\mu \rightarrow \nu_\tau,t \right) & = & \left\{\frac{4 X^2}{(1\!+\!
X^2)^2}\! - \!\frac{64\sqrt{3}X(1\!-\!X^2)}{29(1\!+\!X^2)^2}\,\frac{\alpha^{(e)
}\!-\!\beta^{(e)}}{m_\tau}\cos\left(\varphi^{(\nu)}\! - \!\varphi^{(e)}\right)
\right.  \nonumber \\
& + & \left.O\left[\left(\frac{\alpha^{(e)}\!\!-\!\beta^{(e)}}{m_\tau}\right)^2
\right]\right\}\,\sin^2\left(\frac{m_{\nu_2}^2\!-\!m_{\nu_1}^2}{4 |\vec{p}|
}\,t\right) \nonumber \\
& + & \frac{16 X^2}{841(1+X^2)^2}\left(\frac{\alpha^{(e)}}{m_\mu}\right)^2 
\nonumber \\
& \times & \left\{\sin^2\left[\frac{m_{\nu_2}^2\! - \!m_{\nu_0}^2}{4|\vec{p}|}
\,t\! + \!\frac{1}{2}\left(\arg V_{12} - \varphi^{(\nu)} - 180^\circ + 
\arg V_{11}\right)\right] \right. \nonumber \\
& \; & - \sin^2\!\left.\!\left[\frac{m_{\nu_1}^2\!-\!m_{\nu_0}^2}{4 |\vec{p}|}
\,t\! + \!\frac{1}{2}\left(\arg V_{12} - \varphi^{(\nu)} - 180^\circ + \arg 
V_{11}\right)\right]\right\}\;. \nonumber \\ & &
\end{eqnarray}

\vspace{-0.15cm}

\ni In the third Eq. (16) there appears (in the cubic perturbative order) the 
CP--violating phase

\vspace{-0.35cm}

$$
\arg (V_{10}^*V_{21}^* V_{20} V_{11}) = \arg V_{12} - \varphi^{(\nu)} - 
180^\circ + \arg V_{11} = O\left[\frac{\alpha^{(e)}\!\!-\!\beta^{(e)}}{m_\tau}
\left(\varphi^{(\nu)}\! - \!\varphi^{(e)}\right) \right]\;,
$$

\vspace{-0.13cm}

\ni invariant under any lepton rephasing ({\it cf.} Appendix B). Its form 
presented on the rhs holds in the lepton phasing as in Eq. (13), and can be 
expressed through $\varphi^{(\nu)} - \varphi^{(e)}$ by means of Eqs. (B.5) and 
(B.6). Then, it turns out to be vanishing if $\varphi^{(\nu)} - \varphi^{(e)} 
= 0 $, leading in such a case to a real, CP--preserving matrix $\widehat{V}$ 
in the convenient lepton phasing (B.8). Note that in the lowest (quadratic) 
perturbative order the mass difference $ m^2_{\nu_2} - m^2_{\nu_0}$ is not 
present in the second Eq. (16). 

 The atmospheric neutrino experiments seem to indicate that the $\nu_\mu 
\rightarrow \nu_\tau $ oscillation amplitude is of the order $ O(1) $ [5,6,7].
So, let us take as an input for the leading oscillation amplitude in the third 
Eq. (16) the reasonable value

\vspace{-0.15cm}

\begin{equation}
\frac{4 X^2}{(1 + X^2)^2} \sim 0.5 \;\;{\rm or}\;\; 0.8 \;,
\end{equation}

\vspace{-0.15cm}

\ni where the second figure(or one a bit larger) is more reliable. This gives 
$ X \sim \sqrt{2} - 1 = 0.414 $ or $ (\sqrt{5} - 1)/2 = 0.618 $, and then from 
Eq. (10)

\vspace{-0.15cm}

\begin{equation}
20.3\,\frac{\mu^{(\nu)}}{\beta^{(\nu)}} \sim 1 \;\;{\rm or}\;\;0.5 \;.
\end{equation}

\ni Thus, $ \beta^{(\nu)}/\mu^{(\nu)} \sim 20.3 $ or 40.6 and $ \mu^{(\nu)}/
\beta^{(\nu)} \sim 0.0493 $ or 0.0246.

 As another input let us accept the recent Super--Kamiokande bound [6,7]

\vspace{-0.15cm}

\begin{equation}
|m_{\nu_2}^2 - m_{\nu_1}^2| \sim (0.03\;\;{\rm to}\;\;1)\times 10^{-2}\;
{\rm eV}^2
\end{equation}

\vspace{-0.1cm}

\ni with the preferable value $ 0.5\times 10^{-2}$ eV$^2 $.

 Making use of Eqs. (8) we get
 
\vspace{-0.20cm}

\begin{eqnarray}
m_{\nu_2}^2 - m_{\nu_1}^2 & = & 2 |M_{12}^{(\nu)}| \left( M_{11}^{(\nu)} + 
M_{22}^{(\nu)}\right) \left[1 + \left( \frac{M_{11}^{(\nu)} - M_{22}^{(\nu)}}
{2 |M_{12}^{(\nu)} |}\right)^2 \right]^{1/2} \nonumber \\ 
& = &  20.9\, \beta^{(\nu)}\mu^{(\nu)} \sqrt{1 + \left(20.3 \frac{\mu^{(
\nu)}}{\beta^{(\nu)}} \right)^2 } \nonumber \\  & \sim & (600\;\;{\rm or}
\;\;949)\mu^{(\nu)\,2}\;,
\end{eqnarray}

\vspace{-0.15cm}

\ni where Eq. (18) is used. From Eqs. (19) and (20) we infer that
 
\vspace{-0.2cm}

\begin{equation}
\mu^{(\nu)} \sim (0.707\;\;{\rm to}\;\;4.08)\times 10^{-3}\;
{\rm eV~~~or~~}\;(0.562\;\;{\rm to}\;\;3.25)\times 10^{-3}\;{\rm eV}\;,
\end{equation}

\vspace{-0.1cm}

\ni and then from Eq. (18)
 
\vspace{-0.2cm}
 
\begin{equation}
\beta^{(\nu)} \sim (0.144\;\;{\rm to}\;\;0.828)\times 10^{-1}\;{\rm eV~~~or~~}
\;(0.228\;\;{\rm to}\;\;1.32)\times 10^{-1}\;{\rm eV}\;.
\end{equation}

\vspace{-0.1cm}

\ni With the values (18) and (21), the neutrino mass formulae (8) predict

\vspace{-0.2cm}

\begin{eqnarray}
m_{\nu_1} & \sim & -2.82 \mu^{(\nu)} = -(0.199\;\;{\rm to}\;\;1.15)\times 
10^{-2}\;{\rm eV} \nonumber \\
& {\rm or} & -10.8 \mu^{(\nu)} = -(0.607\;\;{\rm to}\;\;3.51)\times 10^{-2}
\;{\rm eV} \;,\nonumber \\
m_{\nu_2} & \sim & 24.6 \mu^{(\nu)} = (0.174\;\;{\rm to}\;\;1.00)\times 10^{-1}
\;{\rm eV}\nonumber \\
& {\rm or} & 32.6 \mu^{(\nu)} = (0.183\;\;{\rm to}\;\;1.06)\times 10^{-1}
\;{\rm eV} 
\end{eqnarray}

\vspace{-0.1cm}

\ni and, of course,~$ m_{\nu_0} \ll m_{\nu_1}$~if $\varepsilon^{(\nu)\,2}\simeq
0 $ is small enough. The minus sign at~$m_{\nu_1}$ in Eq. (23) is irrelevant in
the relativistic dynamics ({\it cf.} the Dirac equation) and so, can be changed
into the plus sign, if the mass~~$m_{\nu_1}$~~is considered from the phenomeno%
logical point of view.

 The neutrino mass $m_{\nu_1}$ as estimated in Eq. (23) leads to

\vspace{-0.2cm}

\begin{equation}
|m_{\nu_1}^2 - m_{\nu_0}^2| \sim (0.04\;\;{\rm to}\;\;1)\times 10^{-4}\;
{\rm eV}^2\;\;\;{\rm or}\;\;\;(0.04\;\;{\rm to}\;\;1)\times 10^{-3}\;{\rm eV}^2
\;,
\end{equation}

\vspace{-0.1cm}

\ni where $ m_{\nu_0}^2 = \mu^{(\nu)\,2}\varepsilon^{(\nu)\,4}/841 \simeq 0 $ 
is neglected. Such an estimate may be used to evaluate the $\nu_e \rightarrow 
\nu_\mu $ and $\nu_e \rightarrow \nu_\tau $ oscillation probabilities (in the 
vacuum) from the first and second Eq. (16), where the oscillation amplitudes 
are

\vspace{-0.2cm}

\begin{eqnarray}
\frac{16}{841(1+X^2)} \left(\frac{\alpha^{(e)}}{m_\mu}\right)^2 & \sim & 
2.2^{+3.3}_{-2.2} \times 10^{-4}\;\;\;{\rm or}\;\;\; 1.8^{+2.7}_{-1.8}
\times 10^{-4}\;, \nonumber \\
\frac{16 X^2}{841(1+X^2)} \left(\frac{\alpha^{(e)}}{m_\mu}\right)^2 & \sim & 
3.7^{+5.5}_{-3.7}\times 10^{-5}\;\;\;{\rm or}\;\;\;7.0^{+10.5}_{-7.0}
\times 10^{-5}\;.
\end{eqnarray}

\vspace{-0.1cm}

 According to recent estimations [8], the familiar two--flavor neutrino--%
oscillation formula (in the vacuum),

\vspace{-0.25cm}

$$
P = \sin^2 2\theta \sin^2\left(\frac{\Delta m^2}{4|\vec{p}|}\,t\right)\;,
$$

\vspace{-0.1cm}

\ni requires the oscillation amplitude

\vspace{-0.25cm}

$$
\sin^2 2\theta \sim 0.65 \;\;{\rm to}\;\; 1
$$

\vspace{-0.1cm}

\ni and the (unrealistic?) mass--squared difference

\vspace{-0.2cm}

$$
\Delta m^2 \sim (5 \;\;{\rm to}\;\;8)\times 10^{-11}\,{\rm eV}^2
$$

\vspace{-0.1cm}

\ni in order to explain the solar--neutrino deficit. Thus, our amplitudes (25)
of $\nu_e \rightarrow \nu_\mu $ and $\nu_e \rightarrow \nu_\tau$ oscillations
(in the vacuum) are much too small, while the related neutrino mass--squared 
difference (24) is much too large.

According to the recent estimations [8], the two--flavor neutrino oscillations,
strengthened by the resonant MSW mechanism in the Sun matter [9], may solve the
problem of solar neutrinos, if

\vspace{-0.25cm}

$$
(\sin^2 2\theta)_{\rm MSW} \sim 8\times 10^{-3}\;\;,\;\;\Delta m^2 \sim 5
\times 10^{-6}\,{\rm eV}^2
$$

\vspace{-0.1cm}

\ni (the preferred small--mixing--angle solution) or

\vspace{-0.2cm}

$$
(\sin^2 2\theta)_{\rm MSW} \sim 0.6\;\;,\;\;\Delta m^2 \sim 1.6 \times 
10^{-5}\,{\rm eV}^2
$$


\ni (the alternative large--mixing--angle solution). We can see that our 
neutrino mass--squared difference (23) is formally not inconsistent with both
MSW solutions, favouring the second.

 In the present paper there is left open the actual question about interpretat%
ion of LSND events from Los Alamos [10]. They suggest the existence of $\nu_\mu
\rightarrow \nu_e $ oscillations with $\Delta m^2 $ of one to two orders of 
magnitude larger than the Super--Kamiokande $\Delta m^2 $ for atmospheric--\-%
neutrino events. Evidently, the LSND events are relevant for the problem of
existence of only three conservative neutrinos. In fact, they seem to suggest 
the existence of one extra (sterile) neutrino ({\it cf. e.g.} Ref. [11]; for a 
possible origin of the hypothetic sterile neutrino {\it cf.} the end of 
Appendix~A).

\vspace{0.3cm}

\ni {\bf 4. Up and down quarks}

\vspace{0.3cm}

 In the case of up and down quarks we will assume, similarly as for charged 
leptons, that the off--diagonal elements of the mass matrices $\widehat{M}^{(u,
d)} = \left(M^{(u,d)}_{ij}\right)\;\;(i,j = 0,1,2)$ described in Eq. (1) can be
considered as small perturbations of the diagonal elements (this assumption 
will be verified {\it a posteriori}, when we estimate the coupling constants
$\alpha^{(u,d)}/\mu^{(u,d)}$ and $\beta^{(u,d)}/\mu^{(u,d)}$. Then, in the
lowest (quadratic) perturbative order we get

\vspace{-0.2cm}

\begin{eqnarray}
m_{u,d} & = & \frac{\mu^{(u,d)} }{29} \left[\varepsilon^{(u,d)\,2} - \frac{36}{
320 - 5\varepsilon^{(u,d)\,2} }\left(\frac{\alpha^{(u,d)}}{\mu^{(u,d)}}
\right)^2 \right]\; , \nonumber \\ 
m_{c,s} & = & \frac{\mu^{(u,d)}}{29} \left[\frac{4}{9}\left(80 + 
\varepsilon^{(u,d)\,2}\right) + \frac{36}{320 - 5\varepsilon^{(u,d)\,2} }\left(
\frac{\alpha^{(u,d)} }{\mu^{(u,d)} }\right)^2\right. \nonumber \\ 
& - & \left.\frac{10800}{31696 + 1350 C^{(u,d)} + 29\varepsilon^{(u,d)\,2} }
\left(\frac{\alpha^{(u,d)} - \beta^{(u,d)} }{\mu^{(u,d)} }\right)^2\right] \; ,
\nonumber \\ 
m_{t,b} & = & \left.\frac{\mu^{(u,d)}}{29}\right.\left[\frac{24}{25}\left(624 +
25 C^{(u,d)} + \varepsilon^{(u,d)\,2}\right)\right. \nonumber \\ 
& + & \left.\frac{10800}{31696 + 1350 C^{(u,d)} + 29\varepsilon^{(u,d)\,2}}
\left(\frac{\alpha^{(u,d)} - \beta^{(u,d)}}{\mu^{(u,d)}}
\right)^2 \right] \;,
\end{eqnarray}

\ni where $C^{(d)} = 0$ due to Eq. (2).

 These mass formulae imply the relations analogical to Eqs. (4) for charged 
leptons, namely

\vspace{-0.2cm}

\begin{eqnarray}
m_{t,b} & = & \frac{6}{125} \left(351 m_{c,s} - 136 m_{u,d} \right) + \frac{
\mu^{(u,d)}}{29}\,24 C^{(u,d)} \nonumber \\ & - & \frac{105192}{3625} 
\frac{\alpha^{(u,d)\,2}/\mu^{(u,d)}}{320 - 5\varepsilon^{(u,d)\,2}} + 
\frac{24094800}{3625} \frac{\left(\alpha^{(u,d)} - \beta^{(u,d)}\right)^2/
\mu^{(u,d)}}{31696 + 1350 C^{(u,d)} + 29\varepsilon^{(u,d)\,2}}\;,
\nonumber \\
\varepsilon^{(u,d)\,2} & = & \frac{320m_{u,d}}{9 m_{c,s} - 4m_{u,d}} + 
O\left[\left(\frac{\alpha^{(u,d)}}{\mu^{(u,d)}}\right)^2 \right] + O\left[
\left(\frac{\alpha^{(u,d)} - \beta^{(u,d)}}{\mu^{(u,d)}}\right)^2 \right]\;, 
\nonumber \\ 
\mu^{(u,d)} & = & \frac{29}{320}\left(9 m_{c,s} - 4m_{u,d}\right) +
O\left[\left(\frac{\alpha^{(u,d)}}{\mu^{(u,d)}}\right)^2 \right]\mu^{(u,d)} 
+ O\left[\left(\frac{\alpha^{(u,d)} - \beta^{(u,d)}}
{\mu^{(u,d)}}\right)^2 \right]\mu^{(u,d)} \;. \nonumber \\ & &
\end{eqnarray}

\vspace{-0.1cm}

 The mass $ m_s $ may be predicted from the first Eq. (27) in the zero perturb%
ative order, if $ m_d $ and $ m_b $ are known. When the small ratio $ m_d/m_s $
is of the order of relative perturbative corrections [{\it cf.} Eq. (26)], it 
can be also neglected, and then

\vspace{-0.1cm}

\begin{equation}
m_s \simeq \frac{125}{6\cdot 351} m_b = 267\; {\rm MeV}
\end{equation}

\vspace{-0.1cm}

\ni for $m_b = 4.5 $ GeV as an input. Similarly, the third Eq. (27) or, 
equivalently, the second Eq. (26) gives in the zero perturbative order

\vspace{-0.1cm}

\begin{equation}
\mu^{(d)} \simeq \frac{29\cdot 9}{320} m_s \simeq 218\; {\rm MeV}\;,
\end{equation}

\vspace{-0.1cm}

\ni while, due to the estimation (41) of $(\alpha^{(d)}/\mu^{(d)})^2$ discussed
later on, the first Eq.(26) leads jointly with its perturbation to

\vspace{-0.1cm}

\begin{equation}
\varepsilon^{(d)\,2} \simeq \frac{29m_d}{\mu^{(d)}} + \frac{36}{320}
\left(\frac{\alpha^{(d)}}{\mu^{(d)}} \right)^2 \simeq 0.932 + 1.41 = 2.34
\end{equation}

\vspace{-0.1cm}

\ni for $ m_d = 7 $ MeV as another input. If, however, the first Eq. (26) gives
$ m_d < 0 $, then with the input $|m_d| = 7 $ MeV one obtains $\varepsilon^{(d)
\,2} \simeq -0.932 + 1.41 = 0.478 $. Here, $C^{(d)} = 0 $. If $C^{(d)} > 0 $, 
then $ m_s $ is smaller. For instance, $m_s \simeq 200 $ MeV and $\mu^{(d)} 
\simeq 163 $ MeV, when $ C^{(d)} \simeq 8.37\, $.

 In an analogical way,

\vspace{-0.2cm}

\begin{equation}
\mu^{(u)} \simeq \frac{29\cdot 9}{320} m_c = 1060\; {\rm MeV}
\end{equation}

\vspace{-0.1cm}

\ni and

\vspace{-0.1cm}
							
\begin{equation}
\varepsilon^{(u)\,2} \simeq \frac{29m_u}{\mu^{(u)}} + \frac{36}{320}
\left(\frac{\alpha^{(u)}}{\mu^{(u)}} \right)^2 \simeq 0.109 + 0.239 = 0.348
\end{equation}


\ni for $ m_u = 4 $ MeV and $ m_c = 1.3 $ GeV as inputs. However, if $ m_u < 0 
$, then with the input $|m_u| = 4 $ MeV one gets $\varepsilon^{(u)\,2} \simeq 
-0.109 + 0.239 = 0.130 $. Further, from the first Eq. (27)

\vspace{-0.1cm}

\begin{equation} 
C^{(u)} \simeq \frac{29}{24}\,\frac{1}{\mu^{(u)}}\left(m_t - \frac{6\cdot 
351}{125}\,m_c \right) \simeq 175
\end{equation}


\ni for $ m_t = 175 $ GeV as another input. If the constant $ C^{(u)}$ were 
known from some conjecture, the value of one of the masses, $ m_c $ or $ m_t $,
could be a prediction.

 The unitary matrices $\widehat{U}^{(u,d)}$, diagonalizing the mass matrices
$\widehat{M}^{(u,d)}$ through the relations $\widehat{U}^{(u,d)\,\dagger}
\widehat{M}^{(u,d)}\widehat{U}^{(u,d)} = $ diag($m_{u,d}\,,\,m_{c,s}\,,\,
m_{t,b}$), take in the lowest (linear or quadratic) perturbative order the form

\vspace{-0.2cm}

\begin{eqnarray}
& & \widehat{U}^{(f)} = \left(\begin{array}{ccc} A^{(f)}_0 & 0 & 0 \\
0 & A^{(f)}_1 & 0 \\ 0 & 0 & A^{(f)}_2 \end{array}\right) \nonumber \\
& & \nonumber \\ & & + \frac{1}{29}\left(\begin{array}{ccc} 0 & 2\frac{
\alpha^{(f)}}{m_{c,s}}e^{i\varphi^{(f)}} & \frac{16\sqrt{3}}{29}\frac{\alpha^{
(f)}(\alpha^{(f)} - \beta^{(f)})}{m^2_{t,b}}\,e^{2i\varphi^{(f)}} \\ - 2\frac{
\alpha^{(f)}}{m_{c,s}}e^{-i\varphi^{(f)}} & 0 & 8\sqrt{3}\frac{\alpha^{(f)} - 
\beta^{(f)}}{m_{t,b}}e^{i\varphi^{(f)}} \\ \frac{16\sqrt{3}}{29}\frac{\alpha^{
(f)}(\alpha^{(f)} - \beta^{(f)})}{m_{c,s} m_{t,b}}\,e^{-2i\varphi^{(f)}} & 
-8\sqrt{3}\frac{\alpha^{(f)} - \beta^{(f)}}{m_{t,b}}e^{-i\varphi^{(f)}} & 0 
\end{array} \right)\,,\nonumber
\\ & &
\end{eqnarray}

\ni where $ (f) = (u,d) $, while 

\vspace{-0.2cm}

\begin{eqnarray}
A^{(u,d)}_0 & = & 1 - \frac{1}{2}\,\frac{4}{841}\left(\frac{\alpha^{(u,d)}}{
m_{c,s}} \right)^2 \;, \nonumber \\
A^{(u,d)}_1 & = & 1 - \frac{1}{2}\,\frac{4}{841}\left(\frac{\alpha^{(u,d)}}{
m_{c,s}} \right)^2 - \frac{1}{2}\,\frac{192}{841}\left(\frac{
\alpha^{(u,d)} - \beta^{(u,d)}}{m_{t,b}} \right)^2 \;, \nonumber \\ 
A^{(u,d)}_2 & = & 1 - \frac{1}{2}\,\frac{192}{841}\left(\frac{
\alpha^{(u,d)} - \beta^{(u,d)}}{m_{t,b}} \right)^2 \;.
\end{eqnarray}

\vspace{-0.1cm}

\ni Here, in the mass denominators, we keep only leading terms.

 The down--quark weak--interaction states $ d^{({\rm w})}\,,\,s^{({\rm w})}\,,
\,b^{({\rm w})} $  are related to their experimentally observed mas states $ 
d\,,\,s\,,\,b $ by  the unitary transformation

\vspace{-0.2cm}

\begin{equation}
\left(\begin{array}{c} d^{({\rm w})} \\ s^{({\rm w})} \\ b^{({\rm w})} 
\end{array}\right) = 
\widehat{V}\left(\begin{array}{c} d \\ s \\ b \end{array}\right)\;.
\end{equation}

\vspace{-0.1cm}

\ni Here,

\vspace{-0.2cm}

\begin{equation}
\widehat{V} = \widehat{U}^{(u)\,\dagger}\widehat{U}^{(d)}
\end{equation}

\vspace{-0.1cm}

\ni is the familiar (quark) \CKM matrix [to be distinguished from its lepton
counterpart (12)]. Using both Eqs. (34), we can calculate $\widehat{V} = \left(
V_{ij} \right)\;\;(i,j = 0,1,2)$ from the formulae $ V_{ij} = \sum_{k}U^{(u)*}
_{ki}U^{(d)}_{kj} $. In the lowest (linear or quadratic) perturbative order in
$\alpha^{(u,d)}/\mu^{(u,d)}$ and $\beta^{(u,d)}/\mu^{(u,d)}$, we obtain

\vspace{-0.2cm}

\begin{eqnarray}
V_{us}\; \left(\right.\!\! & \!\!\equiv\!\! & V_{01} \left.\right)\; = \; - 
V^*_{cd} \left(\; \equiv \;\; - V^*_{10}\right)  =  \frac{2}{29}
\left(\frac{\alpha^{(d)}}{m_s} e^{i\varphi^{(d)}} - \frac{\alpha^{(u)}}{m_c} 
e^{i\varphi^{(u)}} \right) \;,\nonumber \\ 
V_{cb}\; \left(\right.\!\! & \!\!\equiv\!\! & V_{12}\left.\right)\; = \; -
V^*_{ts} \left(\;\equiv \; - V^*_{21} \right)  =  \frac{8\sqrt{3}}{29}\left( 
\frac{\alpha^{(d)} - \beta^{(d)}}{m_b} e^{i\varphi^{(d)}} - \frac{\alpha^{(u)} 
- \beta^{(u)}}{m_t} e^{i\varphi^{(u)}} \right) 
\nonumber \\
& \!\! \simeq \!\!& \;\frac{8\sqrt{3}}{29}\frac{\alpha^{(d)} - \beta^{(d)}}
{m_b} e^{i\varphi^{(d)}} \;,\nonumber \\
V_{ub}\; \left(\right.\!\! & \!\!\equiv\!\! & V_{02}\left.\right)\; = \; - 
\frac{16\sqrt{3}}{841}\left[ \frac{\alpha^{(u)}(\alpha^{(d)} - \beta^{(d)}
)}{m_c m_b} e^{i(\varphi^{(u)}+\varphi^{(d)})} - \frac{\alpha^{(u)}
(\alpha^{(u)} - \beta^{(u)})}{m_c m_t} e^{2i\varphi^{(u)}} \right. 
\nonumber \\ 
&\!\! - \!\!& \; \left.\frac{\alpha^{(d)}(\alpha^{(d)} - \beta^{(d)})}
{m^2_b} e^{2i \varphi^{(d)}} \right] \simeq - \frac{16\sqrt{3}}{841}\, 
\frac{\alpha^{(u)}(\alpha^{(d)} - \beta^{(d)})}{m_c m_b} e^{i(\varphi^{(u)} + 
\varphi^{(d)})} \;, \nonumber \\  
V_{td}\; \left(\right.\!\! & \!\!\equiv\!\! & V_{20}\; \left.\right)\; = \; -
\frac{16\sqrt{3}}{841} \left[ \frac{(\alpha^{(u)} - \beta^{(u)})\alpha^{(d)}}
{m_t m_s} e^{-i(\varphi^{(u)}+\varphi^{(d)})} - \frac{\alpha^{(d)}(\alpha^{(d)}
- \beta^{(d)})}{m_s m_b}\,e^{-2i\varphi^{(d)}} \right. \nonumber \\ 
& \!\!-\!\! & \left.\frac{\alpha^{(u)}(\alpha^{(u)} - \beta^{(u)})}{m^2_t}\,
e^{-2i\varphi^{(u)}}\right] \simeq \frac{16\sqrt{3}}{841}\frac{\alpha^{(d)}
(\alpha^{(d)}- \beta^{(d)})}{m_s m_b}\,e^{-2i\varphi^{(d)}} \;,
\nonumber \\
V_{ud}\;\left(\right.\!\! & \!\!\equiv\!\! & V_{00}\left.\right)\;\simeq 
\;|V_{ud}|\,,\; V_{cs}\;\left(\right.\equiv \,V_{11}\left.\right)\;\simeq 
\;|V_{cs}|\,,\; V_{tb}\;\left(\right.\equiv \,V_{22}\left.\right)\;\simeq 
\;|V_{tb}|\,.
\end{eqnarray}

\vspace{-0.1cm}

\ni The approximate equalities in Eqs. (38) are due to $ m_t \gg m_b $ and $ 
m_b > m_c $, and also to $\alpha^{(u)} = 2\alpha^{(d)}$ and $\beta^{(u)} = 
\beta^{(d)} \simeq \alpha^{(d)}/2 $ (this gives $\alpha^{(u)} m_b/\alpha^{(d)}
m_c = 6.3 $).

 Taking the experimental value $ |V_{cb}| = 0.041 \pm 0.003 $ as an input, we 
calculate from the second Eq. (38)

\vspace{-0.2cm}

\begin{equation}
\alpha^{(d)} - \beta^{(d)} \simeq \frac{29}{8\sqrt{3}} m_b |V_{cb}| = 
(386 \pm 28)\;\; {\rm MeV}
\end{equation}

\vspace{-0.1cm}

\ni for the $ m_b = 4.5 $ GeV already used in Eq. (28). Hence, invoking Eq. 
(2), we evaluate 

\vspace{-0.2cm}

\begin{eqnarray}
\alpha^{(d)} = 2(\alpha^{(d)} - \beta^{(d)}) \simeq (772 \pm 56)\;\;{\rm MeV}\;
,\nonumber \\
\alpha^{(u)} = 2 \alpha^{(d)} \simeq (1544 \pm 112)\;\;{\rm MeV}\;,
\nonumber \\
\beta^{(u)} = \beta^{(d)} \simeq \frac{1}{2}\alpha^{(d)} \simeq (386 \pm 28)
\;\; {\rm MeV}\;.
\end{eqnarray}

\vspace{-0.1cm}

\ni We can see from Eqs. (30), (32) and (40) that

\begin{eqnarray}
\left(\frac{\alpha^{(u)}}{\mu^{(u)}}\right)^2 & \simeq & 2.12 = O(1)\;\;,\;\; 
\left(\frac{\alpha^{(d)}}{\mu^{(d)}}\right)^2\, \simeq \,12.5 = O(10)\;\;,
\nonumber \\ 
\left(\frac{\alpha^{(u)} - \beta^{(u)}}{\mu^{(u)}}\right)^2 & \simeq & 1.19 = 
O(1)\;\;,\;\; \left(\frac{\alpha^{(d)} - \beta^{(d)}}{\mu^{(d)}}\right)^2 \,
\simeq \, 3.14 = O(1)\;\;.
\end{eqnarray}

\ni In spite of these nonperturbative values, the small relative numerical 
coefficients in the formulae (26) for $ m_{c,s} $ and $ m_{t,b} $ cause that 
the effective perturbative corrections are small (maximally of 1\%, in the case
of $ m_{s} $). But, for $ m_{u,d} $ the perturbative effects are essential [as
large as 220\% and 150\%, respectively; {\it cf.} Eqs. (32) and (30)], what, 
strictly speaking, invalidates the perturbative calculation in this case (where
this calculation ought to be replaced by the numerical evaluation of $ m_{u,d} 
$).

 The value of $\alpha^{(u)}$ as estimated in Eqs. (40) leads through the second
and third Eq. (38) to the prediction

\begin{equation}
\left|\frac{V_{ub}}{V_{cb}}\right| \simeq \frac{2}{29}\frac{\alpha^{(u)}}{m_c} 
\simeq 0.082 \pm 0.006
\end{equation}

\ni for the $m_c = 1.3$ GeV already used in Eq. (32) [the corrections following
from the neglected terms in $ V_{ub}$ do not change the figure (42)]. Hence, we
predict equivalently $|V_{ub}| \simeq 0.0034 \pm 0.0005\,$. We can see that the
result (42) agrees neatly with the experimental value $|V_{ub}/V_{cb}| = 0.08
\pm 0.02 $ [4].

 With $\alpha^{(u)}$ and $\alpha^{(d)}$ as given in Eqs. (40), we find from 
the first Eq. (38) that

\begin{equation}
V_{us} \simeq \frac{2}{29}\,\frac{772 \pm 56}{267}\left[1 - \frac{2\cdot267}
{1300}e^{i(\varphi^{(u)} - \varphi^{(d)})} \right]e^{i\varphi^{(d)}}\;,
\end{equation}

\ni and then

\begin{equation}
|V_{us}|^2 \simeq \left(\frac{2}{29}\,\frac{772 \pm 56}{267}\right)^2
\left[1 + \left(\frac{534}{1300}\right)^2 - \frac{2\cdot534}{1300}\cos\left(
\varphi^{(u)} - \varphi^{(d)}\right) \right]\;.
\end{equation}

\ni Hence, taking the experimental value $|V_{us}| = 0.2205 \pm 0.0018 $ as
another input, we evaluate $\cos(\varphi^{(u)}-\varphi^{(d)}) \simeq -0.0658 $
and

\begin{equation}
\varphi^{(u)} - \varphi^{(d)} \simeq -86.2^\circ + 180^\circ \;,
\end{equation}

\ni when the central values are used. Then, from Eq. (43) we calculate $\tan
(\arg V_{us} - \varphi^{(d)}) \simeq -0.399 $ and

\begin{equation}
\arg V_{us} \simeq -21.8^\circ + \varphi^{(d)}\;\;,\;\; \arg V_{cd} \simeq 
21.8^\circ - \varphi^{(d)} - 180^\circ\;.
\end{equation}

\ni From other Eqs. (38) we can see that

\begin{eqnarray}
\arg V_{cb} & \simeq & \varphi^{(d)}\;\;,\;\; \arg V_{ts} \simeq - 
\varphi^{(d)} + 180^\circ\;,\nonumber \\  
\arg V_{ub} & \simeq & \varphi^{(u)} + \varphi^{(d)} - 180^\circ\;\;,\;\; 
\arg V_{td} \simeq - 2\varphi^{(d)} \;,\nonumber \\  
\arg V_{ud} & \simeq & \arg V_{cs} \simeq \arg V_{tb} \simeq 0 \;.
\end{eqnarray}

 Further, making use of Eqs. (46) and (47) with (45), we predict two mutually 
dependent CP--violating phases:

$$
\arg (V_{us}^*\,V_{cb}^*\, V_{ub} \,V_{cs}) \simeq 21.8^\circ + \varphi^{(u)} -
\varphi^{(d)} - 180^\circ \simeq -64.4^\circ
$$

\ni and 

$$
\arg (V_{cd}^*\,V_{ts}^*\, V_{td} \,V_{cs}) \simeq -21.8^\circ\;.
$$

\ni These, being invariant under any quark rephasing, reduce to

\begin{equation}
\arg V_{ub} \simeq -64.4^\circ \;\;,\;\;\arg V_{td} \simeq -21.8^\circ
\end{equation}

\ni in the special quark phasing, where

\begin{eqnarray}
\arg\,V_{ud} = \,\;\;\;\;0\;\; & \!\!, & \!\arg\,V_{us} = \;\;\;0\;\;,\;\;\; 
\arg\,V_{ub} \simeq \,- 64.4^\circ\,, \nonumber \\ 
\arg\,V_{cd} = \,180^\circ\; & \!\!,& \!\arg\,V_{cs} = \;\;\;0\;\;,\;\:\;\arg\,
V_{cb} = \;\;\;\;\;\;\;\,0\;\;, \nonumber \\ 
\arg\,V_{td} \simeq \!-21.8^\circ\!\!\! & \!\!, & \!\arg\,V_{ts} = 180^\circ\,,
\;\;\,\arg\,V_{tb} = \;\;\;\;\;\;\;\,0\;\;. 
\end{eqnarray} 

 Finally, we can evaluate the rest of magnitudes $|V_{ij}|$ of the \CKM matrix
elements (of them $|V_{us}|$ and $|V_{cb}|$ are used as inputs, while $|V_{ub}
|$ was predicted). In fact, we obtain from Eqs. (38)

\begin{eqnarray}
|V_{cd}| & = & |V_{us}| = 0.221 \;\;,\nonumber \\ |V_{ts}| & = & |V_{cb}| = 
0.041 \;\;,\nonumber \\ |V_{td}| & \simeq & \frac{\alpha^{(d)}}{\alpha^{(u)}}
\frac{m_c}{m_s} |V_{ub}| \simeq 0.0083 \;\;,
\end{eqnarray}

\ni and from the perturbative unitarity of $\widehat{V}$

\begin{eqnarray}
|V_{ud}| & \simeq & 1 - \frac{1}{2} |V_{us}|^2 = 0.976 \;\;,\nonumber \\ 
|V_{cs}| & \simeq & 1 - \frac{1}{2} |V_{us}|^2 - \frac{1}{2} |V_{cb}|^2 = 
0.975 \;\;,\nonumber \\ |V_{tb}| & \simeq & 1 - \frac{1}{2} |V_{cb}|^2 
= 0.999 \;\;. 
\end{eqnarray}

 Thus, in the case of convenient phasing (49), we predict the following 
approximate form of \CKM matrix:

\begin{equation}
\widehat{V} \simeq \left(\begin{array}{ccc}
0.976 & 0.221 & 0.0034\,e^{-i\,64^\circ}\\ -0.221 & 0.975 & 0.041 \\  
0.0083\,e^{-i\,22^\circ} & -0.041 & 0.999 \end{array}\right)\;.
\end{equation}

\ni Here, $|V_{us}|$, $|V_{cb}|$, $|V_{ub}|$ and arg$ V_{ub}$ can be considered
as independent. As inputs we used the experimental values of $|V_{us}|$ and $|
V_{cb}|$ as well as $ m_u $, $ m_d $, $ m_c $, $ m_b $ and $ m_t $. We 
predicted  $|V_{ub}|$ and arg$ V_{ub}$ as well as $ m_s $. From these 7 inputs 
we were also able to determine consistently 7 of all 7 + 1 independent 
parameters involved in the mass matrices $\widehat{M}^{(u,d)}$ (only the 
unphysical phase $\varphi^{(u)} + \varphi^{(d)}$ remained undetermined).

 It is interesting to compare our perturbative form (38) of $\widehat{V}$ with
its convenient Wolfenstein parametrization ({\it cf. e.g.} Ref. [12]),

\begin{equation}
\widehat{V} = \left(\begin{array}{ccc}
1 - \lambda^2/2 & \lambda & A\,\lambda^3(\rho - i \eta)\\ 
-\lambda & 1 - \lambda^2/2  & A\,\lambda^2  \\  
A\,\lambda^3(1 - \rho - i\eta) & -A\lambda^2 & 1 \end{array}\right) 
+ 0(\lambda^4)\;\;,
\end{equation}

\ni being the base for the discussion of popular unitary triangle in the 
complex $\rho + i\,\eta $ plane:

\begin{equation}
V_{ub}^* + V_{td} = A\,\lambda^3 + O(\lambda^5) \simeq A\,\lambda^3\;.
\end{equation}

\ni This parametrization can be considered as an expansion in $\lambda $ of the
standard parametrization [4], where

\vspace{-0.2cm}

\begin{eqnarray}
V_{us} - O(\lambda^7) = s_{12}\, \equiv \,\lambda\;\;  & , & V_{cd} - 
O(\lambda^8) = s_{23} \equiv A\,\lambda^2\;\;,\nonumber \\
V_{ub} = s_{13}\, e^{-i\delta} & \!\equiv\! & A\,\lambda^3(\rho - i\eta)
\end{eqnarray}

\vspace{-0.1cm}

\ni and $ c_{ij} = \sqrt{1 - s^2_{ij}}\;\;(s_{ij} > 0 $ and $ c_{ij} > 0 $). Note
that

\vspace{-0.2cm}

$$
V_{ts} = - A\lambda^2 + O(\lambda^4)\;\;{\rm and}\;\;V_{td} = A\lambda^3(1 - 
\rho - i\eta) + O(\lambda^5)\;.
$$


 When considering Eqs. (53) and (52), we get

 
\begin{eqnarray}
\lambda \simeq 0.221\;\; & , &\;\;A \simeq 0.839 \;\;, \nonumber \\ 
\rho \simeq 0.164\;\; & , & \;\;\eta \simeq 0.337 \;\;, \nonumber \\  
\delta \equiv \arctan \frac{\eta}{\rho} \simeq 64^\circ & , & \beta \equiv 
\arctan \frac{\eta}{1-\rho} \simeq 22^\circ \;\; .
\end{eqnarray}


\ni In the unitary triangle (54), $\arg V^*_{ub} = \delta \equiv \gamma $ and 
$\arg V_{td} \simeq - \beta $ if $ O(\lambda^5) $ is neglected in $ V_{td}$ [{
\it i.e.}, on the rhs of Eq. (54)]. According to Ref. [10], the present uncert%
ainties of $\gamma $ and $\beta $ are


\begin{equation}
41^\circ < \gamma < 134^\circ \;\;,\;\;11^\circ < \beta < 27^\circ \;\;.
\end{equation}


\ni Our predictions (56) are consistent with these limits (however, in the 
future, our $\gamma $ and $\beta $ may lie at the new lower and upper exper%
imental limit, respectively).

\vspace{0.3cm}

\ni {\bf 5. Summary}

\vspace{0.3cm}

 We proposed here the unified form (1) of mass matrix for all fundamental 
fermions: neutrinos, charged leptons, up quarks and down quarks. In this frame%
work, their spectral differences are related only to the differences in values 
of the parameters involved, subject to the tentative constraints (2).

 With some inputs, we obtained a number of predictions neatly consistent with 
available experimental data.

 In the case of charged leptons $ e^-\,,\,\mu^-\,,\,\tau^- $, from the inputs 
of $ m_e $ and $ m_\mu $, we predicted $ m_\tau = 1776.80 $ MeV $ + \Delta 
m_\tau $ with $\Delta m_\tau$ denoting a perturbative correction, quadratic in 
coupling constants, which measured the relative strength of the off--diagonal 
part of mass matrix {\it versus} its diagonal part. If the experimental value 
of $m_\tau $ was also taken as an input, then 3 of all 4 independent parameters
in the charged--lepton mass matrix were consistently determined (only the phase
$\varphi^{(e)}$ remained undetermined). This enabled us to evaluate (up to our 
ignorance of the phase $\varphi^{(e)} $) the charged--lepton contribution 
to the lepton \CKM matrix.

 In the case of neutrinos~~$ \nu_e\,,\;\nu_\mu\,,\;\nu_\tau\, $,~~from the~~at%
mospheric--neutrino~~inputs of $|m_{\nu_2}^2-m_{\nu_1}^2| \sim (0.0003 $ to 
$0.01)$ eV$^2 $ and {\it the $\nu_\mu \rightarrow \nu_\tau $ oscillation 
amplitude} $\sim 0.8 $, we predicted $m_{\nu_0} \ll m_{\nu_1} \sim (0.6 $ to $ 
4)\times 10^{-2}$ eV and $m_{\nu_2} \sim (0.2 $ to $1)\times 10^{-1}$ eV. We 
were also able to evaluate (up to our ignorance of the phase $\varphi^{(\nu)}
$) the neutrino contribution to the lepton \CKM matrix and, taking into account
the previous charged--lepton contribution, the whole matrix (up to an unknown 
phase, dependent on $\varphi^{(\nu)} - \varphi^{(e)}$). Then, the neutrino 
oscillations  $\nu_e \rightarrow \nu_\mu $ and $\nu_e \rightarrow \nu_\tau $ 
(in the vacuum) got amplitudes $\sim 2^{+3}_{-2} \times 10^{-4} $ and $ 7^{+11}
_{-7}\times 10^{-5} $, respectively, and $|m_{\nu_1}^2 - m_{\nu_0}^2| \sim 
(0.04 $ to $ 1)\times 10^{-3}\;\,{\rm eV}^2 $. So, while not fitted at all to 
the (unrealistic) vacuum solution of the solar--neutrino problem, these oscil%
lations require to be strengthened by the resonant MSW mechanism in the Sun 
matter to solve this problem.

 In both lepton cases, the number of inputs was 5, and it was sufficient to 
determine consistently 5 of all 7 + 1 independent parameters appearing in the 
charged--lepton and neutrino mass matrices (the very small $\varepsilon^{(\nu)
\,2} \simeq 0 $, the phase $\varphi^{(\nu)} - \varphi^{(e)}$ and the unphys%
ical phase $\varphi^{(\nu)} + \varphi^{(e)}$ remained undetermined).

 In the case of up and down quarks, from the inputs of $ m_u\,,\,m_d\,,\,m_c =
1.3$ GeV , $ m_b = 4.5 $ GeV and $ m_t $ as well as $|V_{us}|$ and $|V_{cb}|
$, we predicted $ m_s \simeq 270 $ MeV as well as $|V_{ub}/V_{cs}| \simeq 
0.082 $ and $\arg V_{ub} \simeq -64^\circ $. Hence, we were able to evaluate 
all elements of the \CKM matrix.

 In both quark cases the number of inputs was 7, and it was sufficient to 
determine consistently 7 of all 7 + 1 independent parameters, involved in the 
up--quark and down--quark mass matrices (only the unphysical phase $\varphi^{(
u)} + \varphi^{(d)}$ remained undetermined).

 I am much indebted to Dr. Danuta Kie{\l}czewska for her advice about recent 
Super--Kamiokande data.


\vspace{0.8cm}

{\centerline{\bf Appendix A: Unified "texture dynamics"}}

\vspace{0.3cm}

\appendix\setcounter{equation}{0}

 Let us introduce the following $ 3\times 3 $ matrices in the space of three 
fermion families: 

$$
\widehat{a} = \left(\begin{array}{ccc} 0 & 1 & 0 \\ 0 & 0 & \sqrt{2} \\ 0 & 0 
& 0 \end{array} \right)\;\;,\;\;\widehat{a}^\dagger = \left(\begin{array}{ccc} 
0 & 0 & 0 \\ 1 & 0 & 0 \\ 0 & \sqrt{2} & 0 \end{array} \right)\;\;.
\eqno({\rm A}.1)
$$

\ni With the matrix

$$
\widehat{n} = \widehat{a}^\dagger\widehat{a} = \left(\begin{array}{ccc} 
0 & 0 & 0 \\ 0 & 1 & 0 \\ 0 & 0 & 2 \end{array} \right)\;\;, \eqno({\rm A}.2)
$$

\ni they satisfy the commutation relations

$$
[\widehat{a}\,,\,\widehat{n}] = \widehat{a}\;,\;[\widehat{a}^\dagger\,,
\,\widehat{n}] = -\widehat{a}^\dagger \eqno({\rm A}.3)
$$

\ni characteristic for annihilation and creation matrices, while $\widehat{n}$ 
plays the role of an occup\-ation--\-number matrix. However, in addition, they 
obey the "truncation" identities

$$
\widehat{a}^3 = 0\,,\, \widehat{a}^{\dagger\,3} = 0\,. \eqno({\rm A}.4)
$$

\ni Note that due to Eqs. (A.4) the bosonic canonical commutation relation
$[\widehat{a}\,,\,\widehat{a}^\dagger] = \widehat{1}$ does not hold, being 
replaced by the relation $[\widehat{a}\,,\,\widehat{a^\dagger}] = $ diag~$(1\,,
\,1\,,\,-2)$.

 In consequence of Eqs. (A.1), (A.2) and (A.3), we get $\widehat{n}|n\rangle = 
n|n\rangle $ as well as $\widehat{a}|n\rangle = \sqrt{n}|n-1\rangle $
and $\widehat{a}^\dagger|n\rangle = \sqrt{n+1}|n+1\rangle\;\;(n = 0,1,2) $,
however, $\widehat{a}^\dagger|2\rangle = 0 $ ({\it i.e.}, $|3\rangle = 0 $) in
addition to $\widehat{a}^\dagger|0\rangle = 0 $ ({\it i.e.}, $|-1\rangle = 0$).
Evidently, $n = 0,1,2 $ may play the role of a vector index in our three--%
dimensional matrix calculus.

 It is natural to expect that the Gell--Mann matrices (generating the horizon%
tal SU(3) algebra) can be built up from $\widehat{a}$ and $\widehat{a}^\dagger
$. In fact,

\begin{eqnarray*}
\widehat{\lambda}_1 & = & \left(\begin{array}{rrr} 0 & 1 & 0 \\ 1 & 0 & 0 \\
0 & 0 & 0 \end{array}\right) = \frac{1}{2}\left(\widehat{a}^2\widehat{a}^{
\dagger} + \widehat{a}\widehat{a}^{\dagger\,2} \right)\;,\\
\widehat{\lambda}_2 & = & \left(\begin{array}{rrr} 0 & -i & 0 \\ i & 0 & 0 \\
0 & 0 & 0 \end{array}\right) = \frac{1}{2i}\left(\widehat{a}^2\widehat{a}^{
\dagger} - \widehat{a}\widehat{a}^{\dagger\,2} \right)\;,\\
\widehat{\lambda}_3 & = & \left(\begin{array}{rrr} 1 & 0 & 0 \\ 0 & -1 & 0  \\
0 & 0 & 0 \end{array}\right) = \frac{1}{2}\left(\widehat{a}^2\widehat{a}^{
\dagger\,2} - \widehat{a}\widehat{a}^{\dagger\,2}\widehat{a} \right)\;,\\
\widehat{\lambda}_4 & = & \left(\begin{array}{rrr} 0 & 0 & 1 \\ 0 & 0 & 0 \\
1 & 0 & 0 \end{array}\right) = \frac{1}{\sqrt{2}}\left(\widehat{a}^2 + 
\widehat{a}^{\dagger\,2} \right)\;,\\
\widehat{\lambda}_5 & = & \left(\begin{array}{rrr} 0 & 0 & -i \\ 0 & 0 & 0 \\
i & 0 & 0 \end{array}\right) = \frac{1}{i\,\sqrt{2}}\left(\widehat{a}^2 - 
\widehat{a}^{\dagger\,2} \right)\;,\\
\widehat{\lambda}_6 & = & \left(\begin{array}{rrr} 0 & 0 & 0 \\ 0 & 0 & 1 \\
0 & 1 & 0 \end{array}\right) = \frac{1}{\sqrt{2}}\left(\widehat{a}^\dagger
\widehat{a}^2 + \widehat{a}^{\dagger \,2}\widehat{a}\right)\;,\\
\widehat{\lambda}_7 & = & \left(\begin{array}{rrr} 0 & 0 & 0 \\ 0 & 0 & -i \\
0 & i & 0 \end{array}\right) = \frac{1}{i\sqrt{2}}\left(\widehat{a}^\dagger
\widehat{a}^2 - \widehat{a}^{\dagger \,2}\widehat{a} \right)\;,\\
\widehat{\lambda}_8 & = & \frac{1}{\sqrt{3}}\,\left(\begin{array}{rrr} 1 & 0 & 
0 \\ 0 & 1 & 0 \\ 0 & 0 & -2 \end{array}\right) = \frac{1}{\sqrt{3}}\left(
\widehat{a}\widehat{a}^\dagger - \widehat{a}^\dagger\widehat{a} \right)\,,
\\
\widehat{1} & = & \left(\begin{array}{rrr} 1 & 0 & 0 \\ 0 & 1 & 0 \\ 0 & 0 & 1
\end{array}\right) = \frac{1}{2}\left(\widehat{a}^2\widehat{a}^{\dagger\,2} 
+ \widehat{a}\widehat{a}^{\dagger\,2}\widehat{a} + \widehat{a}^{\dagger\,2}
\widehat{a}^2\right)\;.
\end{eqnarray*}

\vspace{-1.92cm}

\begin{flushright}
(A.5)
\end{flushright}

\vspace{0.6cm}
\ni Inversely, $\widehat{a} = (\widehat{\lambda}_1 + i\widehat{\lambda}_2)/2 +
\sqrt{2}(\widehat{\lambda}_6+i\widehat{\lambda}_7)/2$ and $\widehat{a}^{\dagger
} = (\widehat{\lambda}_1 - i\widehat{\lambda}_2)/2+ \sqrt{2}(\widehat{\lambda}
_6 - i\widehat{\lambda}_7)/2$. A message we get from these relationships is 
that a horizontal field formalism, always simple (linear) in terms of $ 
\widehat{\lambda}_A\;\;(A = 1,2,\ldots,8)$ and $\widehat{1}$, is generally not 
simple in terms of $\widehat{a}$ and $\widehat{a}^{\dagger}$. In particular, a 
nontrivial SU(3)--symmetric horizontal formalism is not simple in $\widehat{a}$
and $\widehat{a}^{\dagger}$. Inversely, a nontrivial horizontal field formal\-%
ism, if simple (linear and/or quadratic and/or cubic) in terms of $\widehat{a}$
and $\widehat{a}^{\dagger}$, cannot be SU(3)--symmetric.

 Now, let us consider the following ansatz [1]:

$$
\widehat{M}^{(f)} = \widehat{\rho}^{\,1/2}\widehat{h}^{(f)}\widehat{\rho}^{\,
1/2} \;\;\;(f = \nu\,,\,e\,,\,u\,,\,d)\;, \eqno({\rm A}.6)
$$

\ni where

$$ 
\widehat{\rho}^{\,1/2} = \frac{1}{\sqrt{29}}
\left(\begin{array}{rrr} 1 & 0 & 0 \\ 0 & \sqrt{4} & 0 
\\ 0 & 0 & \sqrt{24} \end{array}\right)  \eqno({\rm A}.7)
$$

\ni and 

\begin{eqnarray*}
\widehat{h}^{(f)} & = & \mu^{(f)}\left[\left(1 + 2\widehat{n}\right)^2 
+ \left(\varepsilon^{(f)\,2} -1 \right)\left(1 + 2\widehat{n}\right)^{-2} + 
\widehat{C}^{(f)}\right] \\
& + & \left(\alpha^{(f)}\widehat{1} - \beta^{(f)}
\widehat{n}\right)\widehat{a}e^{i\varphi^{(f)}} + \widehat{a}^\dagger\left(
\alpha^{(f)}\widehat{1} - \beta^{(f)}\widehat{n}\right)e^{-i\varphi^{(f)}}
\end{eqnarray*}    

\vspace{-1.55cm}

\begin{flushright}
(A.8)
\end{flushright}

\ni with $\widehat{n} = \widehat{a}^\dagger\widehat{a}$ and

$$
\widehat{1} + 2 \widehat{n} = \widehat{N} = \left(\begin{array}{rrr} 
1 & 0 & 0 \\ 0 & 3 & 0 \\ 0 & 0 & 5 \end{array}\right)\;\,,\,\;
\widehat{C}^{(f)} =\left(\begin{array}{rrr} 0 & 0 & 0 \\ 0 & 0 & 0 \\ 0 & 0 & 
C^{(f)} \end{array}\right)\;.   \eqno({\rm A}.9)
$$

\ni It is the matter of an easy calculation to show that the matrices (A.6)
get explicitly the form (1).

 In a more detailed construction following from our idea about the origin of 
three fermion families [1], each eigenvalue $ N = 1\,,\,3\,,\,5 $ of the matrix
$\widehat{N}$ corresponds (for any $f = \nu\,,\,e\,,\,u\,,\,d$) to a wave func%
tion carrying $ N = 1\,,\,3\,,\,5 $ Dirac bispinor indices: $\alpha_1,\alpha_2,
\ldots,\alpha_N $ of which one, say $\alpha_1 $, is coupled to the external 
Standard--Model gauge fields, while the remaining $ N-1 = 0\,,\,2\,,\,4\,:\;\,
\alpha_2\,,\ldots,\;\alpha_N $ are fully antisymmetric under permutations. So, 
the latter obey Fermi statistics along with the Pauli principle implying that 
really $ N-1 \leq 4$, because each $\alpha_i = 1,2,3,4 $. Then, the three wave 
functions corresponding to $ N = 1\,,\,3\,,\,5 $ can be reduced to three other 
wave functions carrying only one Dirac bispinor index $\alpha_1$ (and so, spin 
1/2),

\begin{eqnarray*}
\psi^{(f)}_{1\,\alpha_1} & \equiv & \psi^{(f)}_{\alpha_1}\;, \\
\psi^{(f)}_{3\,\alpha_1} & \equiv & \frac{1}{4}\left(C^{-1}\gamma^5 \right)_{
\alpha_2\alpha_3} \psi^{(f)}_{\alpha_1\alpha_2\alpha_3}  = \psi^{(f)}_{\alpha_1
\,12} =  \psi^{(f)}_{\alpha_1\,34}\;, \\
\psi^{(f)}_{5\,\alpha_1} & \equiv & \frac{1}{24}\varepsilon_{\alpha_2\alpha_3
\alpha_4\alpha_5}\psi^{(f)}_{\alpha_1\alpha_2\alpha_3\alpha_4\alpha_5} =
\psi^{(f)}_{\alpha_1\,1234} \;,
\end{eqnarray*} 

\ni and appearing with the multiplicities 1, 4, 24, respectively (the chiral 
representation is used here). In this argument, the requirement of relativistic
covariance of the wave function (and the related probability current) is 
applied explicitly [1]. The weighting matrix $\widehat{\rho}^{\,1/2}$ as given 
in Eq. (A.7) gets as its elements the square roots of these multiplicities, 
normalized in such a way that tr$\,\widehat{\rho} = 1 $.

 Note that all four matrices $\widehat{M}^{(f)}\;\;(f = \nu\,,\,e\,,\,u\,,\,d)$
defined by Eqs. (A.6) --- (A.9) and (A.1) have a common structure, differing 
from each other only by the values of their parameters $\mu^{(f)}$, $
\varepsilon^{(f)\,2}$, $\alpha^{(f)}$, $\beta^{(f)}$, $ C^{(f)}$ and $\varphi^{
(f)}$. We propose the fermion mass matrices to be of this unified form. Then, 
Eqs. (A.6) and (A.8) define a quantum mechanical model for the "texture" of 
mass matrices $\widehat{M}^{(f)}\;\;(f = \nu\,,\,e\,,\,u\,,\,d)$. Such an 
approach may be called "texture dynamics".

 The fermion mass matrix $\widehat{M}^{(f)}$, containing the kernel $\widehat{h
}^{(f)}$ given in Eq. (A.8), consists of a diagonal part proportional to  $
\mu^{(f)}$, and of an off--diagonal part involving linearly $\alpha^{(f)}$ and 
$\beta^{(f)}$. The off--diagonal part of $\widehat{h}^{(f)}$ describes the 
mixing of three eigenvalues

$$
\mu^{(f)}\left[N^2 + \left(\varepsilon^{(f)\,2} - 1\right)N^{-2} + \delta_{N\,
5} C^{(f)}\right]\;\;(N = 1,3,5)  \eqno({\rm A}.10)
$$

\ni of its diagonal part. Beside the term  $\mu^{(f)}C^{(f)}$ that appears only
for $ N = 5 $, each of these eigenvalues is the sum of two terms containing $
N^2 $. They are: ({\it i}) a term  $\mu^{(f)} N^2 $ that may be interpreted as 
an "interaction" of $ N $ elements ("intrinsic partons") treated on the same 
footing, and ({\it ii}) another term

$$
\mu^{(f)}\left(\varepsilon^{(f)\,2} - 1\right)P_N^{2}\;\;{\rm with}\;\;P_N = 
N^{-1} = \left[N!/(N-1)! \right]^{-1} 
$$

\ni that may describe an additional "interaction" with itself of one element 
arbitrarily chosen among~~$ N $~~elements of which the remaining~~$ N-1 $~~are 
undistinguish\-able. Therefore, the total "interaction" with itself of this 
(arbitrarily) distinguished "parton" is~~$\mu^{(f)}[1 + $ $ (\varepsilon^{(f)
\,2} - 1)N^{-2}]$, so it becomes $\mu^{(f)}\varepsilon^{(f)\,2}$ in the first 
fermion family.

 It seems natural to conjecture that each "intrinsic parton" carries a Dirac 
bispinor index. In fact, such a possibility, as already described in the con\-%
text of the weighting matrix (A.7), follows from our idea about the origin of 
three fermion families [1]. Then, for the (arbitrarily) distinguished "parton",
this index, considered in the framework of a fermion wave equation, is coupled 
to the external gauge fields of the Standard Model. Thus, this "parton" carries
a set of Standard--Model charges corresponding to $f = \nu\,,\,e\,,\,u\,,\,d 
$. For the $ N-1 $ undistinguishable "partons", obeying Fermi statistics along 
with the Pauli principle, their Dirac bispinor indices are mutually coupled, 
resulting into Lorentz scalars, while their number $ N-1 = 0,2,4 $ differen\-%
tiates between three fermion families (for each $ f = \nu\,,\,e\,,\,u\,,\,d $).
These "partons" are free of Standard--Model charges.

 Evidently, the intriguing question arises, how to interpret two possible boson
families corresponding to the number $N-1 = 1,3$ of undistinguishable "partons"
[13]. In the present paper this problem is not discussed. Here, we would like 
only to point out that three fermion families $ N = 1,3,5 $ differ from these 
two hypothetic boson families $ N = 2,4 $ by the full pairing of their $ N-1 = 
0,2,4 $ undistinguishable "partons". So, the boson families, containing an odd 
number $ N-1 = 1,3 $ of such "partons", might be considerably heavier.
  
 {\it A priori}, the "intrinsic partons" may be either strictly {\it algebraic}
objects providing fundamental fermions (leptons and quarks) with new family 
degrees of freedom, or may give us a signal of a new {\it spatial} substructure
of fundamental fermions (built up of spatial "intrinsic partons" = preons). Our
idea about the origin of three fermion families [1] chooses the first option. 
The difficult problem of new non--Standard--Model forces, responsible for the 
binding of $ N $ preons within fundamental fermions, does not arise in this 
option.

 However, if the second option is true, then this irksome (though certainly 
profound) problem does arise and must be solved. It seems that in this case the
most natural preon dynamics may be based (at the na\"{\i}ve phenomenological 
level) on a very strong and very shortrange effective attraction $\sum_{ij} 
V^{(N)}_{ij}\;\;(i,j = 1,2,\ldots,N)$ binding $ N $ spin--1/2 preons in some 
S--wave {\it ground} states (and {\it only} in such states). Among these 
preons, one is (arbitrarily) distinguished by carrying a set $ f = \nu\,,\,e\,,
\;u\,,\,d $ of Standard--Model charges, while the remaining $ N-1 $ are undist%
inguishable and obey Fermi statistics along with the Pauli principle. This 
implies (much as in the case of the first option) that $ N-1 \leq 4$ and so, $ 
N-1 = 0,2,4 $ for the halfinteger total spin (that is then 1/2), what is in 
consistency with the phenomenon of three fermion families. In particular, the 
fundamental fermion of the family $ N = 1 $ ({\it i.e.}, the lepton $\nu_e $ or
$ e^- $ or quark $ u $  or $ d $) is essentially nothing else as the (arbitrar%
ily) distinguished preon, but dressed by the Standard--Model radial effects.

In view of Eq. (A.10), this attraction, jointly with the Standard--Model radial
effects, should give

\begin{eqnarray*}
Z^{(f)}\langle N m^{(N)}\!\!\! & \!+\! & \!\!{\it internal}\;\,{\it kinetic}
\;\,{\it contribution} + \sum_{ij}V_{ij}^{(N)}\rangle^{(f)} \\
& \!=\! & \mu^{(f)}\left[N^2 + \left(\varepsilon^{(f)\,2} - 1
\right)N^{-2} + \delta_{N5} C^{(f)}\right] > 0   
\end{eqnarray*}

\vspace{-1.53cm}

\begin{flushright}
(A.11)
\end{flushright}

\ni for the fermion $f = \nu\,,\,e\,,\,u\,,\,d$ from any family $ N = 1,3,5 $. 
Hopefully, the values $\mu^{(\nu)} \sim (0.6\;\;{\rm to}\;\;3)\times 10^{-3}$
eV, $\mu^{(e)} = 85.9924 $ MeV, $\mu^{(u)} \simeq 1060 $ MeV and $\mu^{(d)} 
\simeq 218 $ MeV as well as $ C^{(u)} \simeq 175 $ should be reasonably reprod%
uced in terms of Standard--Model characteristics carried by the (arbitrarily) 
distinguished "parton" within the fundamental fermion $ f $ (in any family $ N
$), as well as in terms of the preon (effective) mass $ m^{(N)}$ and a few new 
parameters introduced through the function $ V^{(N)}_{ij}$.

 Whatever might be the origin of such a phenomenological shortrange attraction,
this would be certainly an exciting physical problem, related or not to the 
(future) quantum gravitation.

 Those of the preons that are free of Standard--Model charges (and play within 
fundamental fermions the role of undistinguishable preons) may form a {\it 
novel} dark matter, transparent for {\it any} Standard--Model interactions
(and their supersymmetric variants). It is so, if they can appear also as {\it 
free} particles and/or as {\it sole} constituents of some new bound states. 
Evidently, such a Standard--Model--dark matter is able to interact gravitation%
ally.

 A characteristic feature of the undistinguishable preons within fundamental 
fermions is that, though they carry no Standard--Model charges, their configur%
ations $ N-1 = 0,2,4 $ corresponding to the families $ N = 1,3,5 $ can mix in 
charge--changing weak interactions. In the case of neutrinos, this implies 
neutrino oscillations (while for charged leptons the mass and weak--interaction
states are identical). 

 One might go a (bold) step further and ask, if a free preon of those carrying 
no Standard--Model charges could participate in some oscillations together 
with the {\it conventional neutrinos} $\nu_e\,,\,\nu_\mu\,,\,\nu_\tau $ (which 
contain one Standard--Model--active preon $ f = \nu $ and $ N-1 = 0,2,4 $ 
Standard--model--sterile preons of the same sort as the considered free preon).
If it could, it would be nothing else as a massive {\it sterile neutrino} of 
Dirac or Majorana type. Note that here the conventional neutrinos $\nu_e\,,\,
\nu_\mu\,,\,\nu_\tau $ are of Dirac type and so, though they are no mass 
states, include both the (active) lefthanded and (sterile) righthanded compon%
ents, the latter coupled only through mass terms ({\it via} the righthanded 
components of mass states). Thus, in this (intriguing) case, the novel dark 
matter would consist just of sterile neutrinos (and/or their bound states), 
although such a sterile neutrino would be an additional one, different from the
sterile $\nu_{e\,R}$ or, more correctly, sterile $\nu_{e\,R} + (\nu_{e\,R})^c $
(to speak of possible mass states, of Majorana type in this example). Here, 
$(\nu_{e\,R})^c = (\nu_e^c)_L \neq \nu_{e\,L}$. Then, when bound within 
fundamental fermions  $f = \nu\,,\,e\,,\,u\,,\,d$ from families $ N = 1,3,5 $, 
such sterile neutrinos would play also the role of $ N -1 = 0,2,4 $ 
undistinguishable preons. 
 
 
\vspace{0.8cm}

{\centerline{\bf Appendix B: Lepton \CKM matrix}}

\vspace{0.3cm}

\appendix\setcounter{equation}{0}

 Let us consider the lepton \CKM matrix $\widehat{V} = (V_{ij})\;\;(i,j = 0,1,
2) $ as given in Eqs. (13) involving four parameters $ X\,,\,\alpha^{(e)}\,,
\;\varphi^{(\nu)}$ and $\varphi^{(e)}$ (if $\beta^{(e)}$ is neglected {\it 
versus} $\alpha^{(e)}$). We are able to ascribe some values only to two of 
these parameters: $ X \sim (\sqrt{5} - 1)/2 = 0.618 $ and $(\alpha^{(e)}/
\mu^{(e)})^2 = 0.020^{+0.029}_{-0.020}$ with $\mu^{(e)} = 85.9924 $ MeV, what 
gives the central value $\alpha^{(e)} = 12 $ MeV [{\it cf.} Eqs. (17) and (6)].
Then,

\begin{eqnarray*} 
\!\!|V_{01}|\!\! & = & \!\!\frac{2}{29}\,\frac{\alpha^{(e)}}{m_\mu}= 0.0079 
\;\;, \\
|V_{10}| & = & \frac{2}{29\sqrt{1+X^2}}\,\frac{\alpha^{(e)}}{m_\mu}
= 0.0057 \;\;, \\
\!\!|V_{12}|\!\! & = & \!\!|V_{21}| = \left[\frac{X^2}{1+X^2}\! + \frac{192}{
871(1+X^2)}\left(\frac{\alpha^{(e)}}{m_\tau}\right)^2\! - \frac{16\sqrt{3}X}{29
\sqrt{1+X^2}}\frac{\alpha^{(e)}}{m_\tau} \cos\left(\varphi^{(\nu)}\! - 
\varphi^{(e)}\right)\right]^{1/2} \\ 
\!\!\!\!& = & \!\!0.53 \;, \\
\!\!|V_{02}|\!\! & = & \!\!0 \;\;,  \\
\!\!|V_{20}|\!\! & = & \!\!\frac{2X}{29\sqrt{1+X^2}}\,\frac{\alpha^{(e)}}{
m_\mu} = 0.0035 \;\;, \\
\!\!|V_{00}|\!\! & = & \!\!
1\;\;, \\
\!\!|V_{11}|\!\! & = & \!\!|V_{22}| = \left[\frac{1}{1+X^2}\! + \frac{192 X^2}{
871(1+X^2)}\left(\frac{\alpha^{(e)}}{m_\tau}\right)^2\! - \frac{16\sqrt{3}X}{29
\sqrt{1+X^2}}\frac{\alpha^{(e)}}{m_\tau} \cos\left(\varphi^{(\nu)}\! - 
\varphi^{(e)}\right)\right]^{1/2} \\ 
\!\!\!\!& = & \!\!0.85 \;\;,
\end{eqnarray*}  

\vspace{-1.5cm}

\begin{flushright}
(B.1)
\end{flushright}

\ni where the numbers correspond to the central value of $\alpha^{(e)}$. Note 
that in the third and seventh Eq. (B.1)

\vspace{-0.25cm}

$$
\frac{16\sqrt{3}X}{29\sqrt{1+X^2}}\frac{\alpha^{(e)}}{m_\tau} \cos\left(\varphi
^{(\nu)} - \varphi^{(e)}\right) = 0.0034\,\cos\left(\varphi^{(\nu)} - \varphi^
{(e)}\right)   \eqno({\rm B}.2)
$$

\vspace{-0.1cm}

\ni and so, it is negligible, giving practically no chance for determining $
\varphi^{(\nu)} - \varphi^{(e)}$ from a (future) experimental value of $|V_{12}
|$. If $\alpha^{(e)} = 0$, then only the elements

\vspace{-0.35cm}

\begin{eqnarray*} 
|V_{12}| & = & |V_{21}| = \frac{X}{\sqrt{1+X^2}} = 0.53\;\;, \\ 
|V_{00}| & = & 1\;\;, \\
|V_{11}| & = & |V_{22}| = \frac{1}{\sqrt{1+X^2}} = 0.85 
\end{eqnarray*}  

\vspace{-1.5cm}

\begin{flushright}
(B.3)
\end{flushright}

\ni remain different from zero.

 In terms of an unknown phase difference $\varphi^{(\nu)} - \varphi^{(e)}$, we 
can evaluate from Eq. (13) the following CP--violating phase:

\vspace{-0.1cm}

$$
\arg(V_{10}^*V_{21}^*V_{20}V_{11}) = (\arg V_{12} - \varphi^{(\nu)} - 180^\circ
) + \arg V_{11}\;, \eqno({\rm B}.4)
$$

\ni where 

\vspace{-0.1cm}

$$
\tan(\arg V_{12} - \varphi^{(\nu)} - 180^\circ) = \frac{8\sqrt{3}(\alpha^{(e)}
- \beta^{(e)})\sin(\varphi^{(\nu)} - \varphi^{(e)})}{29 m_\tau X - 8\sqrt{3}(
\alpha^{(e)} - \beta^{(e)})\cos(\varphi^{(\nu)} - \varphi^{(e)})} 
\eqno({\rm B}.5)
$$

\vspace{-0.1cm}

\ni and

\vspace{-0.1cm}

$$
\tan(\arg V_{11}) = \frac{8\sqrt{3}(\alpha^{(e)} - \beta^{(e)}) X \sin(
\varphi^{(\nu)} - \varphi^{(e)})}{29 m_\tau + 8\sqrt{3}(\alpha^{(e)} - \beta^{(
e)}) X \cos(\varphi^{(\nu)} - \varphi^{(e)})}  \eqno({\rm B}.6)
$$

\vspace{-0.1cm}

\ni depend on $\varphi^{(\nu)} - \varphi^{(e)}$. Since the lhs of (B.4) is 
invariant under any lepton rephasing, it reduces to

\vspace{-0.1cm}

$$
\arg V^{\rm new}_{20} = \arg V_{12} - \varphi^{(\nu)} - 180^\circ \eqno({\rm 
B}.7)
$$

\vspace{-0.1cm}

\ni in the special lepton phasing, where 

\vspace{-0.1cm}

\begin{eqnarray*} 
\arg V^{\rm new}_{00} = \;\;\;\;0  & , & \arg V^{\rm new}_{01}\;\; = \;\;\;\;\;
\;\;0\;\;\;\;\;,\;\; \\ 
\arg V^{\rm new}_{10} = \;\;180^\circ\!\! & , & \arg V^{\rm new}_{11}\;\; = 
\;\arg V_{11}\;\;,\;\;\arg V^{\rm new}_{12}\,\, =\;\;\;\;\;\;0 \;\;\,,\\
& \; & \arg V^{\rm new}_{21}\;\; = \;\;\;\;180^\circ\;\;\,,\;\;\arg V^{\rm 
new}_{22}\;\; = -\arg V_{11}\;.
\end{eqnarray*}  

\vspace{-1.62cm}

\begin{flushright}
(B.8)
\end{flushright}

\ni Here, $\arg V^{\rm new}_{02}$ is irrelevant because of $|V^{\rm new}_{02}| 
= |V_{02}| = 0 $, while $\arg V^{\rm new}_{20}$ is given as in Eq. (B.7).

 Thus, we predict the following approximate form of lepton \CKM matrix:

$$
\widehat{V} \simeq \left(\begin{array}{ccc} 1 & 0.0079 & 0 \\ -0.0057 & 0.85
\exp(i\,\arg V_{11}^{\rm new}) & 0.53 \\ 0.0035 \exp(i\,\arg V_{20}^{\rm 
new}) & -0.53 & 0.85\exp(-i\,\arg V_{11}^{\rm new}) \end{array}\right)
\eqno({\rm B}.9)
$$

\vspace{0.2cm}

\ni with $\arg V_{20}^{\rm new} = \arg V_{12} - \varphi^{(\nu)} - 180^\circ $ 
and $\arg V_{11}^{\rm new} = \arg V_{11}$ expressed in terms of $\varphi^{(\nu)
} - \varphi^{(e)}$ as in Eqs. (B.5) and (B.6). If $\varphi^{(\nu)} - \varphi^{(
e)} = 0 $, then in Eq. (B.9) the CP--violating phases vanish, $\arg V_{20}^{\rm
new} = 0 $ and $\arg V_{11}^{\rm new} = 0 $, what leads to a real matrix $ 
\widehat{V} $. In general, a nonreal matrix $\widehat{V} $ could violate CP--%
parity, but only in processes, where neutrino mass states (instead of their 
weak--interaction states) might be detected experimentally. The expected 
neutrino oscillations are processes, where (nondegenerate) neutrino mass states
may be detected indirectly. There, CP--parity is generally violated [{\it cf.} 
Eq. (16)]. If $\alpha^{(e)} = 0 $, then

$$
\widehat{V} \simeq \left(\begin{array}{ccc} 1 & 0 & 0 \\ 0 & 0.85 & 0.53 \\ 0
& -0.53 & 0.85 \end{array}\right)
\eqno({\rm B}.10)
$$

\ni in the lepton phasing (B.8). With the transformation $\nu^{\rm (m)}_i = 
\sum_j V_{ij} \nu_j $, we can express explicitly the neutrino mass states 
through their weak--interaction states [{\it cf.} Eq. (14)].

 We can see that the lepton matrix $\widehat{V}$ has physically a different 
structure than the quark matrix $\widehat{V}$ evaluated in Eq. (52): the former
gives strong mixing of leptons $\nu_\mu $ and $\nu_\tau $ from the second and
third family, while for the latter rather the quarks $ d $ and $ s $ from the 
first and second family are strongly mixed. This difference, however. follows 
from only quantitative difference in lepton and quark couplings [{\it cf.} Eqs.
(2)] in the unified fermion mass matrix (1).

\vfill\eject

~~~~
\vspace{1.0cm}

{\bf References}

\vspace{1.0cm}

{\everypar={\hangindent=0.5truecm}
\parindent=0pt\frenchspacing

{\everypar={\hangindent=0.5truecm}
\parindent=0pt\frenchspacing

~1.~W.~Kr\'{o}likowski, in {\it Spinors, Twistors, Clifford Algebras and 
Quantum Deformations (Proc. 2nd Max Born Symposium 1992)}, eds. Z.~Oziewicz 
{\it et al.}, 1993, Kluwer Acad. Press; {\it Acta Phys. Pol.}, {\bf B 27}, 
2121 (1996); {\bf B 28}, 1643 (1997); preprint IFT, Warsaw University, 
August 1997 (hep--ph/9709373), to appear in {\it Acta Phys. Pol.}, {\bf B 28} 
[in the last paper, Eqs. (83) and (84) when added to (58) should be corrected 
into the present form (16)].

\vspace{0.2cm}

~2.~H. Georgi and C. Jarlskog, {\it Phys. Lett.} {\bf 86 B}, 297 (1979).

\vspace{0.2cm}

~3.~W. Kr\'{o}likowski, {\it Phys.Rev.} {\bf D 45}, 3222 (1992); {\it Acta 
Phys. Pol.}, {\bf B 23}, 1245 (1992); {\bf B 25}, 1595 (1994).

\vspace{0.2cm}

~4.{\it Review of Particle Physics}, {\it Phys.Rev.} {\bf D 54}, 1 (1996), 
Part I. 

\vspace{0.2cm}

~5.~Y. Suzuki, {\it Neutrino Masses and Oscillations~~---~~Experiment}, in {\it
Proc. of the 28th Inter. Conf. on High Energy Physics}, Warsaw (Poland) 1996, 
eds. Z.~Ajduk and A.K.~Wr\'{o}b\-lewski, World Scientific Pub. 1997; A.Yu.
~Smirnow, {\it Neutrino Masses and Oscillations --- Theory}, in the same {\it 
Proceedings}.

\vspace{0.2cm}

~6.~Y.~Totsuka (Super--Kamiokande Collaboration), in {\it 28th Inter. Symp. on 
Lepton Photon Interactions}, Hamburg (Germany) 1997, to appear in the {\it 
Proceedings}.

\vspace{0.2cm}

~7.~M.~Nakahata (Super--Kamiokande Collaboration), in {\it Inter. Europhysics 
Conf. in High Energy Physics}, Jerusalem (Israel) 1997, to appear in the 
{\it Proceedings}.

\vspace{0.2cm}

~8.~G.L.~~Fogli, E.~~Lisi and D.~~Montanino, preprint~~BARI--TH/284--97 (hep--%
ph /9709473).

\vspace{0.2cm}

~9.~L.~Wolfenstein, {\it Phys.Rev.} {\bf D 17}, 2369 (1978); S.P.~Mikheyev and 
A.Yu.~Smirnow, {\it Sov. Journ. Nucl. Phys.} {\bf 42}, 913 (1986); {\it Nuovo
Cimento}, {\bf C 9}, 17 (1986).

\vspace{0.2cm}

10.~C.~Athanassopoulos {\it et al.} (LSND Collaboration), preprint UCRHEP--%
E97 (nucl--ex/9709006); and references therein.

\vspace{0.2cm}

11.~S.M.~Bilenky, C.~Giunti and W.~Grimus, preprint IASSNS--AST 97/63 + UW Th 
Ph--1997--43 + DFTT 67/97, November 1997 (hep--ph/9711311).

\vspace{0.2cm}

12.~A.J.~Buras, in {\it Symposium on Heavy Flavours}, Santa Barbara (USA) 1997,
to appear in the {\it Proceedings}, preprint TUM--HEP--299/27, October 1997 
(hep--ph/9711217).

\vspace{0.2cm}

13.~W. Kr\'{o}likowski, {\it Phys.Rev.} {\bf D 46}, 5188 (1992); {\it Acta 
Phys. Pol.}, {\bf B 24}, 1149 (1993); {\bf B 26}, 1217 (1995); {\it Nuovo
Cimento}, {\bf 107 A}, 69 (1994).

\vfill\eject

\end{document}